\documentclass{article}
\usepackage[utf8]{inputenc}
\usepackage{graphicx}
\graphicspath{ {./images/} }
\usepackage{enumitem}
\usepackage{url}
\usepackage{nameref}

\usepackage[dvipsnames]{xcolor}
\usepackage{hyperref}

\colorlet{mylinkcolor}{violet}
\colorlet{mycitecolor}{YellowOrange}
\colorlet{myurlcolor}{Aquamarine}
\hypersetup{
    breaklinks=true,
	colorlinks=true, 
	linktoc=all,     
    linkcolor=violet,
    filecolor=magenta,      
    urlcolor=cyan,
    citecolor=YellowOrange
}

\usepackage{amsthm}
\theoremstyle{definition}
\newtheorem{definition}{Definition}[section]
\usepackage{geometry}
\newgeometry{vmargin={35mm}, hmargin={35mm,35mm}}

\usepackage{listings}
\usepackage{titlesec}
\usepackage{titletoc}
\usepackage[title,titletoc]{appendix}

\usepackage{caption}
\usepackage{array,tabularx}
\newenvironment{conditions}
  {\par\vspace{\abovedisplayskip}\noindent
   \tabularx{\columnwidth}{>{$}l<{$} @{${}-{}$} >{\raggedright\arraybackslash}X}}
  {\endtabularx\par\vspace{\belowdisplayskip}}
\usepackage[skins]{tcolorbox}
\newtcolorbox{protocolframe}[2][]{%
  enhanced,colback=white,colframe=black,coltitle=black,
  sharp corners,boxrule=0.4pt,
  fonttitle=\itshape,
  attach boxed title to top left={yshift=-0.3\baselineskip-0.4pt,xshift=2mm},
  boxed title style={tile,size=minimal,left=0.5mm,right=0.5mm,
    colback=white,before upper=\strut},
  title=#2,#1
}
\newcommand{\subsubsubsection}[1]{\paragraph{#1}\mbox{}\\}
\title{\bfseries{Zendoo: a zk-SNARK Verifiable Cross-Chain Transfer Protocol Enabling Decoupled and Decentralized Sidechains}}

\author{
    Alberto Garoffolo\\
    \texttt{alberto@horizen.global}\\
    \texttt{Horizen}
    \and
    Dmytro Kaidalov\\
    \texttt{dmytro.kaidalov@iohk.io}\\
    \texttt{IOHK Research}
    \and
    Roman Oliynykov\\
    \texttt{roman.oliynykov@iohk.io}\\
    \texttt{IOHK Research}\\
    \texttt{V.N.Karazin Kharkiv National University}
}

\date{January 2020}

\begin{document}

\maketitle
\thispagestyle{empty}

\section*{ \centering }
\begin{center}
    \textbf{Abstract}
\end{center}
Sidechains are an appealing innovation devised to enable blockchain scalability and extensibility. The basic idea is simple yet powerful: construct a parallel chain -- sidechain -- with desired features, and provide a way to transfer coins between the mainchain and the sidechain. 

In this paper, we introduce \textit{Zendoo}, a construction for Bitcoin-like blockchain systems that allows the creation and communication with sidechains of different types without knowing their internal structure. We consider a parent-child relationship between the mainchain and sidechains, where sidechain nodes directly observe the mainchain while mainchain nodes only observe cryptographically authenticated certificates from sidechain maintainers. We use zk-SNARKs to construct a universal verifiable transfer mechanism that is used by sidechains.

Moreover, we propose a specific sidechain construction, named \textit{Latus}, that can be built on top of this infrastructure, and realizes a decentralized verifiable blockchain system for payments. We leverage the use of recursive composition of zk-SNARKs to generate succinct proofs of sidechain state progression that are used to generate certificates’ validity proofs. This allows the mainchain to efficiently verify all operations performed in the sidechain without knowing any details about those operations.

\newpage
{
  \hypersetup{linkcolor=black}
  \tableofcontents
}
\section{Introduction}

Since the inception of the Bitcoin cryptocurrency in 2008 \cite{N08}, the topic of decentralized ledger technology has received significant attention among experts from various areas. Bitcoin became the first decentralized payment system based on peer-to-peer networking. Its key feature -- the absence of centralized control -- is claimed to be the disruptive innovation that will help build more robust, fair, and transparent financial systems. Bitcoin inspired the appearance of many other systems based on the same principle of decentralization with a variety of different features.

With the increasing use of Bitcoin and similar blockchain systems, their inherent limitations became apparent: limited throughput, increased latency, reduced ability to scale and expand functionality, etc. \cite{C16}. Even more important is that such decentralized systems are challenging to update since there is no single decision-making entity. Even a small protocol change requires a cumbersome process of community agreement, which makes the introduction of new features difficult.

\begin{figure}[htbp]
	\centering
	\includegraphics[trim={0.5cm 11.7cm 4cm 2.55cm}, clip,width=0.8\columnwidth] {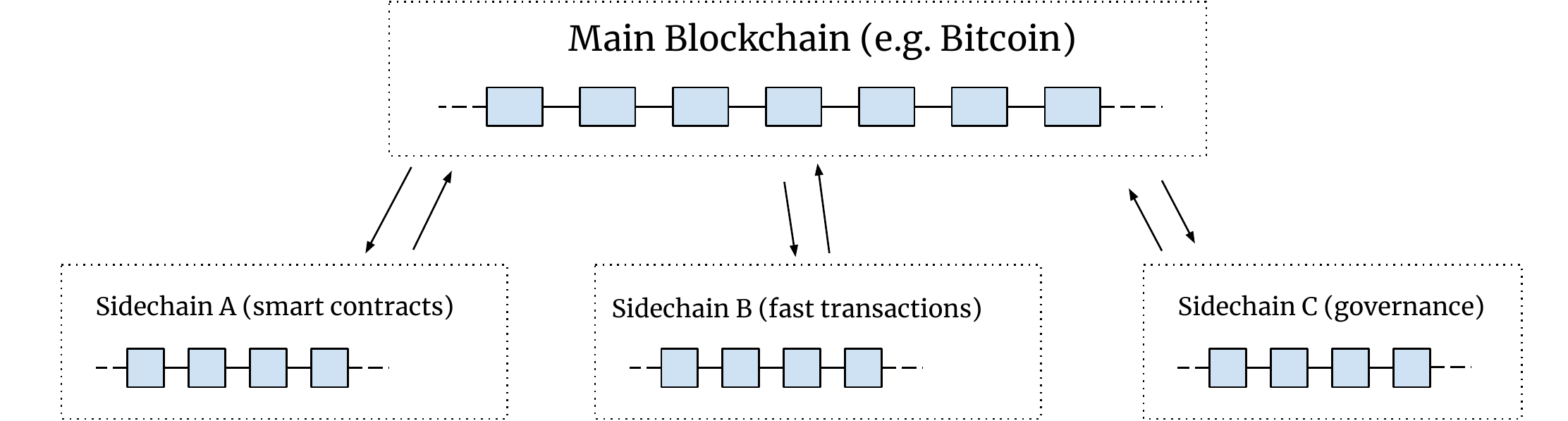}
	\caption{Sidechains. The main blockchain provides basic cryptocurrency functionality while sidechains implement specific functions.}
	\label{fig:1BAsConc}
\end{figure}

Sidechains, proposed by A. Back et.al. in 2014 \cite{BCDF14}, is an appealing concept that allows one to work around the constraints of a single decentralized blockchain. The basic idea is simple: to create a separate blockchain with whatever functionality is needed and provide a way to communicate with the main blockchain (Fig. \ref{fig:1BAsConc}). Communication means the ability to transfer a mainchain native asset (e.g. bitcoins) to and from a sidechain.

This way, for instance, a blockchain system, like Bitcoin, can be extended with additional functionalities (such as smart contracts \cite{RSK}) implemented in a separate sidechain, which uses the same native asset, hence remaining in the Bitcoin ecosystem.

In this paper, we propose \textit{Zendoo}, a universal construction for Bitcoin-like blockchain systems that allows the creation and communication with sidechains of different types without knowing their internal structure (e.g. what consensus protocol is used, what types of transactions are supported, etc.). In fact, the sidechain may not even be a blockchain but can be any system that uses the standardized method to communicate with the mainchain.

Specifically, we consider a parent-child relationship between the mainchain and sidechains, where sidechain nodes directly observe the mainchain while mainchain nodes only observe cryptographically authenticated certificates from sidechain maintainers. Among other things, such certificates authorize transfers coming from sidechains. Certificate authentication and validation are achieved by using zk-SNARKs \cite{BCTV13}, which enable constant-sized proofs of arbitrary computations. The main feature of our construction is that sidechains are allowed to define their own zk-SNARKs, thus establishing their own rules for authentication and validation. The fact that all zk-SNARK proofs comply with the same verification interface used by the mainchain enables great universality as the sidechain can use an arbitrary protocol for authenticating its certificates. E.g., the sidechain may adopt a centralized solution where the zk-SNARK just verifies that a certificate is signed by an authorized entity (like in \cite{BCDF14}) or, for instance, a decentralized chain-of-trust model as in \cite{GKZ19}.

Moreover, we propose a specific sidechain construction, named \textit{Latus}, that can be built on top of this infrastructure, and realizes a decentralized verifiable blockchain system. We leverage the use of recursive composition of zk-SNARKs to generate succinct proofs of sidechain state progression (as in \cite{Coda}) that are used to generate certificate proofs for the mainchain. This allows the mainchain to efficiently verify all operations performed in the sidechain without knowing any details about those operations.
\\~\\
The paper is structured in the following way: section [\ref{sec:2_prelim}] provides basic definitions that are used throughout the paper; section [\ref{sec:3_Overview}] provides a general overview of the sidechain concept; section [\ref{sec:4_Zendoo}] provides details about the proposed cross-chain communication protocol Zendoo and introduces basic interfaces imposed by the mainchain side; section [\ref{sec:5_Latus_Sidechain}] provides details of the Latus sidechain construction.

\subsection{Related Work}

The concept of sidechains was first introduced by A.~Back et.al. in 2014 \cite{BCDF14}. They introduced a general notion of a 2-way peg and described two operational modes -- synchronous and asynchronous -- to implement interactions between pegged chains. The synchronous mode implies that both main and side chains are aware of each other and can verify transfer transactions directly, while the asynchronous mode relies on validators to process transfers.

Notable construction of sidechains was presented in \cite{PS15,SL16} and called Drivechains. It aims to deploy sidechains on top of the Bitcoin network. While forward transfers (from the mainchain to a sidechain) are processed by providing SPV proofs (like the synchronous mode in \cite{BCDF14}), backward transfers rely upon validators. Validators in Drivechains are mainchain miners who observe sidechains and endorse transfers.

The first formal treatment of sidechains was proposed by P.~Gaži, A.~Kiayias, and D.~Zindros in \cite{GKZ19}. In addition, they presented a sidechain construction for proof-of-stake blockchains where sidechain nodes directly observe and confirm forward transfers while backward transfers are confirmed by certifiers chosen among sidechain block forgers.

Our previous proposal on sidechains \cite{GV18} presents a flexible model which allows the construction of different types of sidechains whose internal structures are unknown to the mainchain. It relies on certifiers to confirm backward transfers in the mainchain. Though, in this model certifiers are chosen randomly from a pool of certifiers registered directly in the mainchain.  

Mentioned constructions differ from our current proposal in various aspects, most notably because they either assume that the mainchain observes sidechains directly or relies on some intermediary to confirm transfers. In addition, they do not provide flexibility (except \cite{GV18}), which means that a sidechain construction (e.g. consensus protocol) cannot be chosen freely.  

In \cite{KZ18}, A. Kiayias and D. Zindros proposed implementation of the sidechain protocol for the proof-of-work blockchains based on smart contracts. Another notable sidechain construction that relies on smart contracts is called Plasma and was presented in \cite{Plasma} by J.~Poon and V.~Buterin. On the contrary, our construction does not rely on smart contracts.

One of the main features of the construction presented in this paper is the usage of zk-SNARKs for enabling verifiable cross-chain communication. zk-SNARK has initially been proposed as a zero-knowledge protocol which allows proving possession of some information without revealing it \cite{BCTV13}. However, this technique is suited for more than simply securing information but also for solving scalability issues: it enables succinct constant size proofs of almost arbitrary computations. For instance, using the recursive composition of zk-SNARKs \cite{Coda,Halo} it is possible to construct a succinct proof of state transition virtually for any number of transactions. We were inspired by these techniques while designing our sidechain construction.

A notable sidechain construction that also relies on zk-SNARKs is ZK Rollup \cite{ZKRollup}. It is a layer 2 solution based on smart contracts for scaling transaction throughput in Ethereum~\cite{ETH}. The basic idea is that transactions are carried out off-chain while the information about entailed state transitions together with a zk-SNARK proof of their validity is submitted to the contract. It still requires submission of some limited information about each transaction on-chain to prevent data availability attacks thus limiting scalability. Our construction differs from ZK Rollup in many aspects, most notably because we do not push sidechain transaction data to the mainchain.

There are many other attempts to construct cross-chain transfer mechanisms including the Liquid project \cite{DPWP16}, Polkadot \cite{GW16}, Interledger \cite{TS16}, Cosmos \cite{Cosmos}, and many others. They propose various solutions that are different from our construction.

\section{Preliminaries} \label{sec:2_prelim}

In this section, we introduce definitions of several cryptographic constructions that are used throughout the paper. We defer formal descriptions (especially of the recursive SNARKs composition) for a separate paper and define here only basic notations needed to describe the proposed sidechain construction.

\subsection{Cryptographic Definitions}

\begin{definition}{\textbf{Collision-Resistant Hash Function (CRH).}}
A hash function $H$ is collision-resistant if the probability of finding two different input strings $a$ and $b$ such that $H(a)=H(b)$ is negligible (more formal definition can be found, e.g., in \cite{Old01}).

Whenever we refer to a hash function, we suppose it is collision-resistant.
\end{definition}

\begin{figure}[htbp]
	\centering
	\includegraphics[trim={2cm 12.6cm 7cm 1.5cm}, clip,width=0.9\columnwidth] {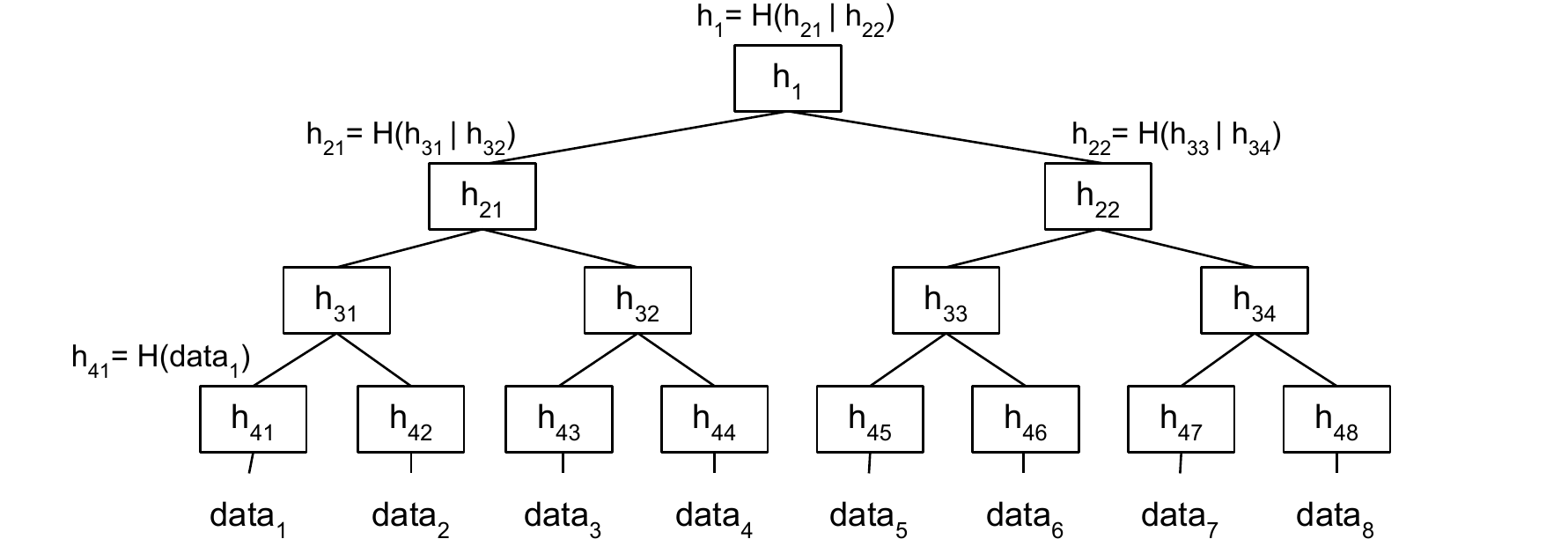}
	\caption{Merkle Hash Tree.}
	\label{fig:MHT}
\end{figure}

\begin{definition}{\textbf{Merkle Hash Tree (MHT).} \label{def:MHT}}
The Merkle Hash Tree, or simply Merkle Tree (MT), is a binary tree data structure where the value of an internal node is computed as the hash of values of its children, and the value of a leaf node is the direct hash of a data block represented by this leaf (see Fig. \ref{fig:MHT}) \cite{M88, NBFMG16}.
\end{definition}

We call the top-level node ($h_1$ in Fig. \ref{fig:MHT}) the root hash of the MHT. Given that a collision-resistant hash function is used to calculate tree nodes, we can consider root hash as a tree authenticator: it is impossible to tamper even a single bit of data in the tree without also changing the root hash.

An important feature of the Merkle tree structure is that it produces a concise proof of a particular data block’s membership in a tree with the particular root hash. E.g., if one wants to prove that $data_4$ (Fig. \ref{fig:MHT}) is included in the MHT tree with the root hash $h_1$, they just need to provide a verifier with the data block along with a tuple of internal nodes $(h_{43}, h_{31}, h_{22})$ that will allow recalculating the tree root and comparing it to the provided root $h_1$. We call it \textbf{Merkle proof}. 

\begin{definition}{\textbf{Succinct Non-Interactive Argument of Knowledge (SNARK).} \label{def:snark}}
\\ A SNARK is a proving system consisting of a triplet of algorithms \textit{(Setup, Prove, Verify)} that allows proving satisfiability of a set of inputs to an arithmetic constraint system (see, e.g., \cite{BCTV13, BG18} for more formal definition and properties analysis).

We define an arithmetic constraint system as a set of polynomials over a finite field $F$ in variables $(x_1,...,x_r,y_1,...,y_s)$. A satisfying assignment for the given constraint system $C$ is an assignment of $F$ elements to $x_i$ and $y_j$ such that all polynomials evaluate to zero. We indicate a satisfying assignment as $C(a,w)$, where $a=(a_1,...,a_r)$, $a_i\in{F}$ and $w=(w_1,...,w_s)$, $w_j\in{F}$. We refer to $a$ as public input and $w$ as witness.

Then, the algorithms $(Setup, Prove, Verify)$ are defined such that
\begin{enumerate}[leftmargin=4em, itemsep=0em]
    \item $(pk,vk) \leftarrow Setup(C,1^\lambda)$ bootstraps SNARK for a constraint system $C$ under security parameter $\lambda$. The bootstrapped SNARK is specified by a pair of keys $(pk,vk)$ which are a proving key and a verification key correspondingly.
    \item $\pi \leftarrow Prove(pk,a,w)$ evaluates a proof $\pi$, which confirms that $(a,w)$ is a satisfying assignment for $C$.
    \item $true/false \leftarrow Verify(vk,a,\pi)$ verifies that $\pi$ is a valid proof attesting to the satisfying assignment $(a,w)$ for the constraint system $C$.
\end{enumerate}

Algorithms $(Setup, Prove, Verify)$ satisfy the following properties:
\begin{enumerate}[leftmargin=4em, itemsep=0em]
    \item \textbf{Completeness}. For any constraint system $C$ and $(a,w)$, if $\pi \leftarrow Prove(pk,a,w)$ is a valid proof, then $Verify(vk,a,\pi)$ is always true.
    \item \textbf{Knowledge soundness}. If a pair $(a,w)$ is not a satisfying assignment for $C$, then the probability of obtaining $\pi$ such that $Verify(vk,a,\pi)=true$ is negligible.
    \item \textbf{Succinctness}. For every constraint system $C$ bootstrapped with $(pk,vk)$ and every $a\in{F^r}$, the size of a proof and verification time is polynomial in $\lambda$.
\end{enumerate}
\end{definition}

\subsection{Recursive SNARKs Composition for State Transitions}

Here, we provide a high-level definition of the recursive proof composition technique that is used in our sidechain model to construct succinct proofs of state transitions. The idea of recursive proofs has been discussed, e.g., in \cite{BCTV13, Coda, Halo}. What follows is based principally on the construction described in \cite{Coda}.

\begin{definition}{\textbf{State Transition System.}}
A state transition system is defined by a set of all possible states $S$, a set of all possible transitions $T$, and a transition function $update(t_i,s_i)$, where $s_i\in{S}$ and $t_i\in{T}$, which returns a new state $s_{i+1}$ or $\bot$ in case $(t_i,s_i)$ does not constitute a valid input for the $update$ function.
\end{definition}

Speaking informally, we would like to define a SNARK that attests to many iterative state transitions. E.g., if we have transitions $(t_1,t_2,...,t_n)$ that are applied sequentially to state $s_1$ to produce state $s_{n+1}$, we would like to have a succinct proof of the following statement: “there exist such $(t_1,...,t_n)$ so that $update(t_n, update(t_{n-1}, update(..., update(t_1,s_1))))=s_{n+1}$”.

By applying this to blockchain, we will be able to provide succinct proofs of transition between some states $s_i$ and $s_j \: (i<j)$. The state can be represented, for instance, as a list of unspent transaction outputs \cite{BitWiki-Tx}, while transitions are regular blockchain transactions that spend some outputs and create new ones. This construction is of great value for verifiable sidechains.

\begin{definition}{\textbf{Recursive SNARKs for state transition systems.} \label{def:2_5_rec_snarks}}
We define recursive SNARKs composition as a tuple of SNARKs $(Base, Merge)$ such that:

\begin{enumerate}
    \item \textbf{Base} is a SNARK for a single transition that proves the existence of such $t$ so that $s_{i+1}=update(t,s_i)$. It is defined by a triplet \textit{(Setup, Prove, Verify)} such that:
    \begin{itemize}
        \item $(pk^{Base},vk^{Base}) \leftarrow Setup(1^\lambda)$ bootstraps \textit{Base} SNARK;
        \item $\pi^{Base} \leftarrow Prove(pk^{Base}, (s_i,s_{i+1}), (t_i))$ evaluates a proof $\pi^{Base}$ that confirms $s_{i+1}=update(t_i, s_i)$;
        \item $true/false \leftarrow Verify(vk^{Base}, (s_i,s_{i+1}), \pi^{Base})$ verifies that $\pi^{Base}$ is a valid proof attesting state transition from $s_i$ to $s_{i+1}$.
    \end{itemize}
    \item \textbf{Merge} is a SNARK that merges two other SNARKs (either \textit{Base} or \textit{Merge}) proving the validity of transition between states $s_i$ and $s_j \: (i+1<j)$. It is defined by a triplet \textit{(Setup, Prove, Verify)} such that:
    \begin{itemize}
        \item $(pk^{Merge},vk^{Merge}) \leftarrow Setup(1^\lambda)$ bootstraps \textit{Merge} SNARK;
        \item $\pi^{Merge} \leftarrow Prove(pk^{Merge}, (s_i,s_j), (s_k,\pi^a_1,\pi^a_2))$ evaluates a proof $\pi^{Merge}$ that confirms $\pi^a_1$, $\pi^a_2$ are valid SNARKs ($a\in{\{Base,Merge\}}$), which attest state transitions from $s_i$ to $s_k$ and from $s_k$ to $s_j$ correspondingly. Altogether, it proves a valid transition from $s_i$ to $s_j \: (i<k<j)$;
        \item $true/false \leftarrow Verify(vk^{Merge}, (s_i,s_j), \pi^{Merge})$ verifies that $\pi^{Merge}$ is a valid proof attesting state transition from $s_i$ to $s_j, i<j$.
    \end{itemize}
\end{enumerate}
\end{definition}

We intentionally omit specifics of the recursive SNARKs composition, which in reality is more sophisticated. More details on the topic can be found, for instance, in \cite{Coda, Halo}. We defer the details of our construction for a separate paper. At this point, we provided only basic definitions, which allow us to describe the sidechain protocol while abstracting away the details of the SNARKs construction.
\section{General Overview} \label{sec:3_Overview}

This section gives an overview of the sidechain concept in general and the main components of any sidechain design. We also briefly discuss our proposed solutions.

Before going any further, we want to introduce abstract definitions of the terms \textit{mainchain} and \textit{sidechain} that are used throughout this paper.

\begin{definition}{\textbf{Mainchain (MC).}\label{def:mc}}
The mainchain is a blockchain system based on the Bitcoin backbone protocol model \cite{GKL14}, which maintains a public ledger of asset-transfer transactions. Additionally, the mainchain supports a standardized mechanism to register and interact with separate sidechain systems. By interaction, we mean the cross-chain transfer protocol, which enables sending a native asset to a sidechain and receiving it back in a secure and verifiable way without the need to know anything about the internal sidechain construction or operations.
\end{definition}

\begin{definition}{\textbf{Sidechain (SC).}\label{def:sc}}
The sidechain is a separate system attached to the mainchain by means of a cross-chain transfer protocol.
\end{definition}

Speaking informally, we consider the mainchain to be a blockchain platform that supports basic payment functionality with some native asset \textit{Coin} (e.g. Bitcoin \cite{N08}, Horizen \cite{H19}, etc.). Then, the sidechain is an attached domain-specific platform that also uses the \textit{Coin} asset (but not limited to it). In our model, we consider a single mainchain with many sidechains attached to it.

The definition of a sidechain (as in Def. \ref{def:sc}) does not imply the usage of any particular data structure or consensus algorithm. The mainchain is totally agnostic to the sidechain construction. It can be another decentralized blockchain, some centralized database maintained by the predefined authority, or more generally, an arbitrary application.

The need to introduce sidechains in a general payment-based blockchain system comes from the need to allow the creation of different blockchain applications that use the same mainchain asset. Creation of such applications directly on the mainchain is not always possible due to inherent technological limitations, such as restricted throughput, expensive storage, etc. Sidechains effectively solve these problems.

\subsection{Main Components of a Sidechain Design}

Analyzing existing attempts to design sidechains \cite{GV18, BCDF14, PS15, SL16, KZ18, GKZ19, ZKRollup}, we may outline three basic components that underlie any sidechain architecture:
\begin{enumerate}[leftmargin=5em, itemsep=0em]
    \item \textbf{Mainchain consensus protocol (MCP)}.
    \item \textbf{Cross-chain transfer protocol (CCTP)}.
    \item \textbf{Sidechain consensus protocol (SCP)}.
\end{enumerate}

Depending on a specific design, these components can be highly coupled with each other or decoupled so that the mainchain is almost independent from any particular sidechain implementation.

In our construction, we aspire to multipurposeness and, thus, designing a system so that the MCP and SCP are completely decoupled. The CCTP is naturally a bridge between them and is unified and fixed by the mainchain consensus protocol. On the other end, the SCP can be freely defined by a sidechain developer. This allows a variety of different sidechains with different purposes to thrive while not requiring any changes to the mainchain.
\\~\\
\textbf{Cross-chain transfer protocol}. The CCTP protocol defines the communication between the mainchain and sidechain(s). Basically, it is a 2-way peg protocol that allows sending coins back and forth. At a high level, it defines two basic operations:
\begin{itemize}[leftmargin=5em, itemsep=0em]
    \item \textbf{Forward Transfer}, and
    \item \textbf{Backward Transfer}.
\end{itemize}

A forward transfer sends coins from the mainchain to a sidechain. A backward transfer, correspondingly, moves coins back from the sidechain to the mainchain. These operations are the cornerstone of overall sidechain construction. The backward transfer is of particular importance since we do not want to oblige the mainchain to track sidechains and, thus, it cannot directly verify the validity of withdrawals coming from them. That is why most of the focus in developing sidechains is directed toward constructing secure and reliable backward transfers.

In our approach, Zendoo, we consider a forward transfer as a special transaction on the mainchain that destroys coins and provides sidechain-specific metadata allowing a user to receive coins in the sidechain. Implementation of forward transfers is straightforward as it does not require the mainchain to know anything about the sidechain state for validation.

A more complex procedure is required for backward transfers. They are initiated in the sidechain as special transactions, batched in a \textit{withdrawal certificate}, and propagated to the mainchain for processing. Since we do not want the MC to follow the SC state -- as this would impose enormous computational and storage burden on the MC and, thus, undermine the whole point of having sidechains -- the question arises how to implement validation of backward transfers in the most efficient and secure way. 

In our previous paper \cite{GV18}, this problem has been addressed by introducing into the system a special type of decentralized actors -- \textit{certifiers} -- that were registering themselves in the MC and were responsible for signing withdrawal certificates. Although the safety of this approach has been shown, it requires certain assumptions about an honest majority of certifiers, which, in some scenarios, may not be the case.

In Zendoo, we avoid direct reliance on certifiers or any other special type of actors assigned to validate withdrawal certificates. Instead, we are going to leverage SNARKs \cite{BCTV13, BG18, Coda} to provide means for the mainchain to effectively validate withdrawals.
\\~\\
\textbf{Sidechain consensus protocol}. We consider the SCP as a generalized notion that encompasses all the details about a particular sidechain construction such as consensus algorithm, accounting system, types of supported transactions, incentives mechanism, a protocol for withdrawal certificate generation, etc. Also, importantly, each sidechain defines its own SNARK\footnote{Speaking more formally, here, we refer to an arithmetic constraint system (arithmetic circuit) that is compiled for each sidechain and defines the logic of the SNARK. Note that the interface of the verifier is defined by the mainchain so -- even though the internal logic of the SNARK may be different for different sidechains -- generated proofs can be verified in the standardized way.} that is used to validate withdrawal certificates. This provides flexibility to define its own rules for backward transfers. For instance, a sidechain can adopt a chain-of-trust model \cite{GKZ19} or even the certifiers model \cite{GV18}. It is completely decoupled from the mainchain consensus protocol, which will just invoke a unified verifier to validate a proof.

Even though the SCP can be designed in different ways, we propose one specific construction of a decentralized verifiable sidechain based on the Ouroboros protocol \cite{KRDO17}. We will call this construction Latus. In short, we are going to use recursive composition of SNARKs to generate succinct proofs of sidechain state transitions. Each withdrawal certificate commits to the SC state whereas the SNARK proof validates transition between states committed by successive withdrawal certificates. Since backward transfers are a part of the sidechain state transition, they are also validated by the proof.
\\~\\
The following section [\ref{sec:4_Zendoo}] introduces Zendoo, a cross-chain communication protocol for sidechains which is principally about the definition of the transfer protocol and how a new sidechain can be registered in the mainchain. It defines the sidechains design from the mainchain point of view. Then, in section [\ref{sec:5_Latus_Sidechain}], we describe in detail the proposed sidechain construction Latus.
\section{Zendoo: a Cross-Chain Transfer Protocol for Sidechains} \label{sec:4_Zendoo}

The following section provides details about the communication protocol between the mainchain and sidechains, which is primarily represented by the cross-chain transfer protocol. We show how the CCTP protocol is integrated in the mainchain, what interfaces are provided, and how a new sidechain can be created.

\subsection{Cross-Chain Transfer Protocol} \label{sec:CCTP}

The cross-chain transfer protocol is the cornerstone of our sidechain design as it connects the mainchain with all sidechains spawned from it. Its main function is to allow sending coins to sidechains and receiving them back in a secure and reliable way. This section provides a high-level specification of forward and backward transfers and how they are integrated into the mainchain.

\subsubsection{Forward Transfers} \label{sec:4_1_1_ForwardTransfers}

The design of forward transfers is straightforward and similar to many existing proposals for sidechains \cite{BCDF14, PS15, SL16, KZ18, GKZ19} as well as to our original proposal \cite{GV18}.

On the mainchain side, it is implemented as a special type of operation (we will call it \textit{Forward Transfer}) that destroys coins and provides metadata for withdrawing coins in a sidechain. Then, it is the responsibility of the sidechain to sync forward transfers from the MC and issue the corresponding amount of coins.

\begin{definition}{\textbf{Forward Transfer (FT).}\label{def:FC}}
Forward Transfer is an operation that moves coins from the original blockchain \textit{A} (the mainchain) to the destination sidechain \textit{B}. It is represented by a tuple of the form:
\[FT\ \stackrel{\mathrm{def}}{=}\ (ledgerId,\ receiverMetadata,\ amount),\]
where:
\begin{conditions}
    ledgerId & a unique identifier of a previously created and active sidechain to which coins are transferred; \\
    amount &  a number of coins to transfer; \\
    receiverMetadata &  some metadata for receiving sidechain \textit{B} (e.g., a receiver’s address); its structure is not fixed in the mainchain and can consist of different variables of predefined types depending on a sidechain’s construction; its semantic meaning is not known to the mainchain.
\end{conditions}
\end{definition}

There can be several approaches to integrate forward transfers on the mainchain side depending on its details. For instance, forward transfer can be a separate transaction type which destroys coins in the mainchain, or, in the case of a UTXO-based blockchain system (e.g. Bitcoin or Horizen), we can consider FT as a special unspendable transaction output in a regular multi-input multi-output transaction \cite{BitWiki-Tx, AA17}.

To be more specific and facilitate further reading, we assume that the mainchain has a UTXO-based accounting model\footnote{It is in line with our own implementation of sidechains, which is going to be deployed on the Horizen mainchain.}. Then, a regular transaction with forward transfers may have the following structure:

\begin{protocolframe}{}
\begin{lstlisting}
type Transaction {
   Inputs: {
      Input(addr: 0x013A.., amount: 5, signature: 0x034B..),
      Input(addr: 0x0930.., amount: 3, signature: 0x1AA1..),
      ...
   }
   Outputs: {
      Output(addr: 0x023B.., amount: 1),
      Output(addr: 0x0732.., amount: 2),
      ForwardTransfer(ledgerId: 0x300C.., receiverInfo: 0x139D.., amount: 2),
      ForwardTransfer(ledgerId: 0x300C.., receiverInfo: 0x893D.., amount: 3),
      ...
   }
}
\end{lstlisting}
\end{protocolframe}{}

Given that an FT is a non-spendable output, it basically destroys coins in the mainchain, and the amount of transferred coins is verified by the mainchain as part of the overall transaction verification.

\subsubsection{Backward Transfers} \label{sec:4_1_2_BackwardTransfers}

The backward transfer protocol allows coins to move from a sidechain to the mainchain, and, as in \cite{GV18} and \cite{GKZ19}, it relies on the idea of batched transfers. This means that all requested backward transfers submitted to the sidechain during a certain period -- called the “withdrawal epoch” -- are collected in a special withdrawal certificate and pushed to the mainchain for processing.

Withdrawal certificates are more than just a container for backward transfers, they are a kind of sidechain heartbeat that is periodically submitted to the mainchain even though there might be no backward transfers\footnote{Later, in section [\ref{sec:5_Latus_Sidechain}], we will show how this can be leveraged to construct secure verifiable sidechains by providing their state commitments as part of withdrawal certificates }.

A withdrawal epoch is defined by a range of MC blocks. Withdrawal epochs for different sidechains are not aligned and may have a different length ($epoch\_len$ parameter is set upon sidechain creation), and, therefore, the entire system runs asynchronously. 

\begin{figure}[htbp]
	\centering
	\includegraphics[trim={1.5cm 11.03cm 4cm 4.35cm}, clip,width=0.9\columnwidth] {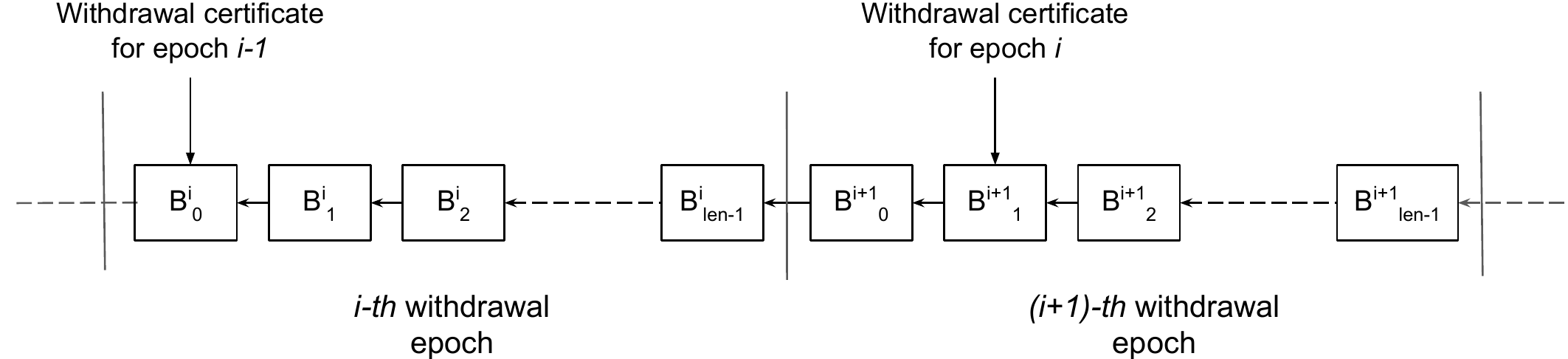}
	\caption{Withdrawal epochs in the mainchain. Note that withdrawal epochs for different sidechains may not overlap. It depends on parameters that have been set upon sidechain creation.}
	\label{fig:4_1}
\end{figure}

Let us define an MC block $B$ that belongs to a specific epoch as $B_j^{ep\_id}$, where $ep\_id$ is the epoch number and $j\in{[0,epoch\_len)}$ is the serial number of the block within the epoch (Fig.~\ref{fig:4_1}).

The sidechain is obliged to submit a withdrawal certificate for epoch $i$ during the first $submit\_len$ blocks of the epoch $i+1$ ($submit\_len$ is a system parameter). If a withdrawal certificate has not been submitted during this time, the sidechain is considered \textit{\textbf{ceased}} (see Def.~\ref{def:ceased_sc}) and no more withdrawal certificates for this sidechain will be accepted by the mainchain (however, the funds can still be withdrawn with a ceased sidechain withdrawal [\ref{sec:mainchain_managed_withdrawals}]).

\begin{definition}{\textbf{Ceased Sidechain.}\label{def:ceased_sc}}
A sidechain is deemed ceased by the mainchain if a withdrawal certificate for that sidechain has not been submitted on time, i.e. a certificate for withdrawal epoch $i$ has not been submitted during the first $submit\_len$ blocks of the epoch $i+1$\footnote{Even though this requirement may seem strong, it is necessary to provide certain properties which we discuss later. We also explore the possibility to provide more flexibility for withdrawal certificate submission}. 
\end{definition}

Note that the mainchain consensus protocol does not impose any rules on how exactly a withdrawal certificate should be generated and by whom it should be submitted. It is up to the sidechain to define corresponding procedures. We only assume that it is submitted by means of a special transaction in the mainchain.

As it has been mentioned, a withdrawal certificate contains backward transfers. We may consider them as requests that are fulfilled once included in the withdrawal certificate and propagated to the mainchain. There are no restrictions for how backward transfers should be submitted and collected (e.g. it can be a separate transaction on the SC side).

\begin{definition}{\textbf{Backward Transfer (BT).}\label{def:backward_transfer}}
Backward Transfer is an operation that moves coins from the sidechain \textit{B} to the original mainchain \textit{A}. It is represented by a tuple of the form:
\[BT\ \stackrel{\mathrm{def}}{=}\ (receiverAddr,\ amount),\]
where:
\begin{conditions}
    receiverAddr & an address in the mainchain where transferred coins should be credited; \\
    amount &  the number of transferred coins.
\end{conditions}
\end{definition}

There can be different approaches to integrate backward transfers in the mainchain. Following the assumption of a UTXO-based mainchain, a BT can be represented by a special output in a transaction with a withdrawal certificate.

\begin{definition}{\textbf{Withdrawal Certificate (WCert).} \label{def:wcert}}
Withdrawal certificate is a standardized posting that allows sidechains to communicate with the mainchain. Its main functions are:

1) delivering backward transfers to the MC; and

2) serving as a heartbeat message enabling the MC to identify SC status.\\
It is represented by a tuple of the form:
\[WCert\ \stackrel{\mathrm{def}}{=}\ (ledgerId,\ epochId,\ quality,\ BTList,\ proofdata,\ proof),\]
where:
\begin{conditions}
    ledgerId & an identifier of the sidechain for which WCert is created; \\
    epochId &  a number of a withdrawal epoch; \\
    quality & an integer value that indicates the quality of this withdrawal certificate (explained later); \\
    BTList & a list of backward transfers included in this withdrawal certificate; \\
    proofdata & input data to a SNARK verifier; \\
    proof & a SNARK proof.
\end{conditions}
\end{definition}

Now, we discuss in more detail the substance of certificate parameters as it is one of the most important parts of the sidechain design.

As it has been briefly outlined, the basis of the proposed construction is that there are no special entities that authorize withdrawal certificates (e.g., like certifiers in \cite{GV18} or slot leaders in \cite{GKZ19}). Instead, the certificate authorization and validation rely completely on the included SNARK proof, and the SNARK itself is defined by the sidechain. The mainchain knows only the verification key -- which is registered upon sidechain creation -- and the interface of the verifier, which is unified for all sidechains. If the SNARK proof and public parameters are valid, then the certificate gets included and processed in the mainchain.
\\~\\
\textbf{Withdrawal certificate quality}. It might happen that several withdrawal certificates appear for the same sidechain in the same withdrawal epoch\footnote{Even though it should not happen under normal operation, it may be the case, for instance, if the sidechain is a blockchain system, which experiences a continuous fork, or due to some malicious activity}. Since only one WCert should be selected among them and given that the mainchain does not know about the sidechain consensus protocol and does not track its state, there should be a mechanism for the mainchain to decide which certificate is the best one. Such a mechanism is realized through the \textit{\textbf{quality}} parameter: the mainchain adopts a certificate with the highest quality or the one that was submitted first in case there are several certificates with equal qualities. The validity of the quality parameter is enforced by the SNARK proof.
\\~\\
\textbf{Withdrawal certificate verification}. Verification of a newly submitted WCert on the mainchain is performed using the following basic rules:
\vspace*{1mm}
\begin{protocolframe}{\textbf{WCert Verification}}
\begin{enumerate}
    \item \textit{\textbf{ledgerId}} should be an identifier of a currently active sidechain;
    \item \textit{\textbf{epochId}} should be a valid withdrawal epoch number for the $ledgerId$ (remember that the certificate should be submitted during the first $submit\_len$ blocks of the epoch following the one, for which such certificate was created);
    \item \textit{\textbf{quality}} should be higher than the quality of the previously submitted withdrawal certificate for this epoch; if it is the first WCert for this epoch - any quality is accepted;
    \item \textit{\textbf{proof}} should be a valid SNARK proof whose verification key $vk_{WCert}$ is set upon sidechain registration;
\end{enumerate}
\end{protocolframe}

SNARK verification is the most essential part of the verification procedure as it encapsulates verification of backward transfers and other parameters provided within the certificate. The basic SNARK verifier interface is the following:

\[ true/false \leftarrow Verify(vk_{WCert},\ public\_input,\ proof),\]
\vspace*{-6mm}
\[ public\_input\ \stackrel{\mathrm{def}}{=}\ (wcert\_sysdata,\ MH(proofdata)),\]
where:
\begin{conditions}
    vk_{WCert} & a SNARK verification key registered upon the sidechain creation; \\
    wcert\_sysdata &  a part of the public input, which is unified for all sidechains and enforced by the mainchain (explained further); \\
    proofdata & a part of the input data that is defined by the sidechain and passed along the withdrawal certificate; it is basically a list of variables of predefined types whose semantics are not known to the mainchain; \\
    MH(proofdata) & a root hash of a Merkle tree where leaves are variables from proofdata; it is essential for the SNARK to keep a list of public inputs short, thus we combine them in a tree and pass the root hash only\footnote{A full payload of \textit{proofdata} is provided during the proof generation as a witness.}; \\
    proof & a SNARK proof itself submitted as a part of the certificate.
\end{conditions}

\textbf{\textit{wcert\_sysdata}} parameter plays an important role from the security standpoint. The idea is to allow the mainchain to verify the proof against some public input parameters that are defined by the protocol. For instance, the \textit{BTList} and \textit{quality} parameters that are part of the certificate must be verified before being used by the mainchain. Another example is the mainchain block hashes of the epoch boundaries that must be verified to guarantee that the proof refers to the current epoch and the active chain.

 \textit{wcert\_sysdata} is represented by the tuple of the following form:
 
 \[ wcert\_sysdata\ \stackrel{\mathrm{def}}{=}\ (quality,\ MH(BTList)),\ H(B_{last}^{i-1}),\ B(_{last}^i)),\]
 where:
 \begin{conditions}
    quality & the quality parameter from the withdrawal certificate; \\
    MH(BTList) &  a root hash of a Merkle tree where leaves are backward transfers from the BTList provided within the certificate; \\
    H(B_{last}^{i-1}) & a block hash of the last mainchain block in the withdrawal epoch $i-1$ (given that the certificate is for the epoch $i$); \\
    H(B_{last}^i) & a block hash of the last mainchain block in the withdrawal epoch $i$.
\end{conditions}

The generalized SNARK verifier provides flexibility to implement different SNARKs for different sidechain models. For instance, one may want to implement the sidechain with a centralized cross-chain transfer protocol where withdrawal certificates are verified by a signature from an authorized entity. Or, conversely, a completely decentralized verifiable sidechain can be constructed as will be discussed in [\ref{sec:5_Latus_Sidechain} \nameref{sec:5_Latus_Sidechain}]). 

Succinct proofs and constant time verification make the overall sidechain design particularly appealing as it does not impose a significant burden for the mainchain.

\subsubsubsection{Mainchain Managed Withdrawals} \label{sec:mainchain_managed_withdrawals}

There might be cases when a user would want to request a backward transfer directly from the mainchain rather than creating a BT in the SC. For instance, it would allow users to withdraw funds in case of a misbehaving (e.g., maliciously controlled sidechain that censors submission of backward transfers) or ceased sidechain. 

Hence, we introduced two additional mechanisms that allow users to make withdrawals directly in the mainchain:

\vbox{%
\begin{enumerate}[leftmargin=5em, itemsep=0em]
    \item \textbf{Backward transfer request (BTR)}, and
    \item \textbf{Ceased sidechain withdrawal (CSW)}.
\end{enumerate}
}

We consider each of them as a special type of transaction. Similar to withdrawal certificates, such operations are secured by SNARK proofs.

The \textbf{BTR} is used to withdraw funds from an active sidechain if for some reason a user cannot create a backward transfer inside the sidechain. The idea is that all BTRs submitted to the mainchain will be synchronized to the sidechain and processed there to verify their legitimacy and include the corresponding backward transfers in the next WCert using the standard flow. Such processing can be enforced by the withdrawal certificate SNARK to force a maliciously controlled sidechain to process user’s withdrawals\footnote{Note that it is up to a sidechain construction to define exactly how BTRs are processed (for example, the Latus sidechain construction [\ref{sec:5_Latus_Sidechain} \nameref{sec:5_Latus_Sidechain}] introduces its own method for enforcing BTRs processing).}. Importantly, the BTR does not lead to a direct coin transfer in the mainchain.

The \textbf{CSW} is used to withdraw funds from ceased sidechains. Since withdrawal certificates are not allowed for ceased sidechains, it becomes the only way to retrieve funds. A valid CSW makes direct payment to the submitter.

\begin{definition}{\textbf{Backward transfer request (BTR).} \label{def:btr}}
The BTR is a generic request for a backward transfer that is submitted on the mainchain. It is represented by the following tuple:
\[BTR\ \stackrel{\mathrm{def}}{=}\ (ledgerId,\ receiver,\ amount,\ nullifier,\ proofdata,\ proof),\]
where:
\begin{conditions}
    ledgerId & an identifier of the sidechain, for which BTR is created; \\
    receiver &  an address of the receiver on the mainchain; \\
    amount & the number of coins to be transferred; \\
    nullifier & a unique identifier of claimed coins; \\
    proofdata & input data to a SNARK verifier; \\
    proof & a SNARK proof.
\end{conditions}
\end{definition}

As in the case with a withdrawal certificate, the SNARK for the BTR is defined by the sidechain and represented by the verification key $vk_{BTR}$, which is set upon sidechain registration.

The syntax of the $proofdata$ and $proof$ are the same as for the withdrawal certificate. The basic interface of the SNARK verifier is the following:

\[ true/false \leftarrow Verify(vk_{BTR},\ public\_input,\ proof),\]
\vspace*{-6mm}
\[ public\_input\ \stackrel{\mathrm{def}}{=}\ (btr\_sysdata,\ MH(proofdata)),\]
where:

$vk_{BTR}$ is a SNARK verification key for the BTR registered upon the sidechain creation;

$btr\_sysdata,\ proofdata,\ MH(\cdot)$, and $proof$ have the same meaning as similar parameters in the withdrawal certificate.

\textit{btr\_sysdata} is defined as:

\[ btr\_sysdata\ \stackrel{\mathrm{def}}{=}\ (H(B_w),\ nullifier,\ receiver,\ amount),\]

where $H(B_w)$ is a block hash of the mainchain block where the latest withdrawal certificate for this sidechain has been submitted.

\begin{definition}{\textbf{Ceased Sidechain Withdrawal (CSW).} \label{def:CSW}}
The CSW is an operation that allows the movement of coins from the ceased sidechain \textit{B} to the original mainchain \textit{A}. It is represented by a tuple of the following form:
\[CSW\ \stackrel{\mathrm{def}}{=}\ (ledgerId,\ receiver,\ amount,\ nullifier,\ proofdata,\ proof),\]
where all parameters have the same meaning as in the case of the BTR.
\end{definition}

As it can be seen, BTR and CSW have the same structure, though conceptually they are different because CSW performs direct payment while BTR does not. The interface of the SNARK verifier for the CSW is completely the same as for the BTR.
\\~\\
Additionally, we discuss the role of nullifiers in both BTR and CSW. In the mainchain, a nullifier is an abstract identifier of claimed coins. The mainchain will not allow the submission of two transactions with the same nullifier. The main reason for having the nullifier is to prevent repeated submission of BTRs or CSWs that try to withdraw the same coins (thus, essentially doing double spend). Since the mainchain does not maintain the sidechain state, at the very least, for ceased sidechains, it requires some abstract double-spend prevention mechanism, which is exactly what is provided by nullifiers.
\\~\\
Note that both BTR and CSW are just complementary operations to allow more flexibility in some subtle use cases or in the case of a malfunctioning sidechain. It is up to the sidechain to define how they are used. For instance, one can omit defining these operations at all (e.g., by setting $vk_{BTR}$ and $vk_{CSW}$ to NULL), thus completely relying on the normal withdrawal procedure through withdrawal certificates.

\subsubsubsection{Withdrawal Safeguard}

The \textbf{safeguard} is a special feature introduced to prevent unlimited withdrawals from a sidechain to the mainchain in the case of the malicious sidechain. The essence of the safeguard function is to maintain the balance of each created sidechain and to prevent withdrawing an amount larger than what was previously transferred to that sidechain. A similar idea was introduced in \cite{GKZ19} and \cite{GV18}.

Implementation of the safeguard feature is simple: for each created sidechain, a special balance variable is maintained by the mainchain. Each forward transfer increases the balance by the transferred number of coins, and each withdrawal certificate, or ceased sidechain withdrawal reduces the balance by the withdrawn amount. The WCert and CSW cannot withdraw more coins than are stored in the sidechain balance.

This feature prevents possible implications of sidechain corruption. It guarantees that only the transferred number of coins can be withdrawn back to the mainchain. Even in the case of total corruption or a maliciously constructed sidechain, an adversary cannot mint coins out of thin air.

\subsubsection{Sidechain Transactions Commitment} \label{sec:4_1_3_ScTxsCommitment}

So far, we defined 4 types of actions (that are either separate transactions or outputs in a regular transaction) that determine cross-chain communication from the mainchain point of view:

\vbox{%
\begin{enumerate}[leftmargin=5em, itemsep=0em]
    \item Forward Transfer (FT).
    \item Withdrawal Certificate (WCert).
    \item Backward Transfer Request (BTR).
    \item Ceased sidechain withdrawal (CSW).
\end{enumerate}
}

\begin{figure}[htbp]
	\centering
	\includegraphics[trim={0.5cm 8.6cm 8cm 1.6cm}, clip,width=0.9\columnwidth] {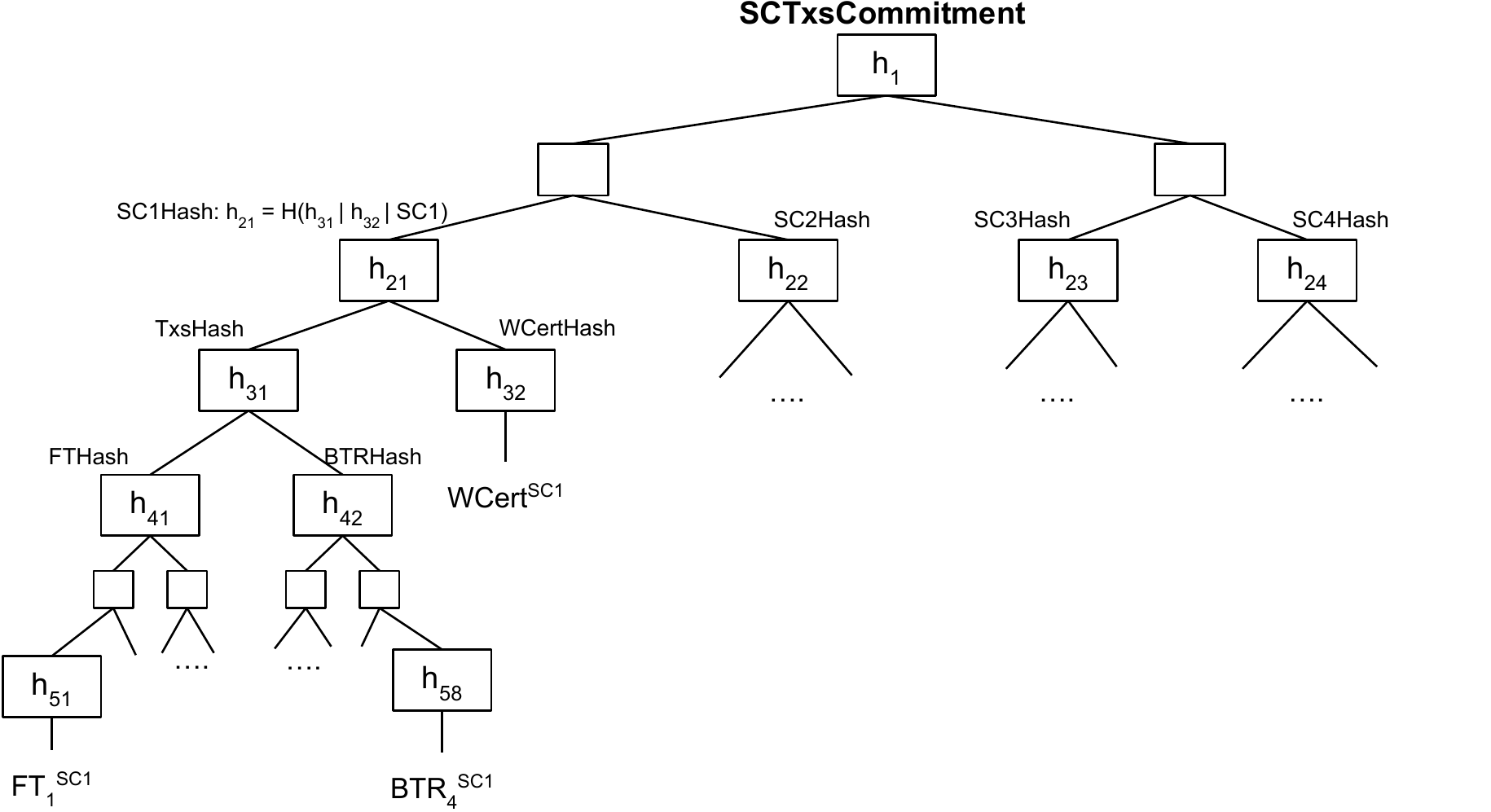}
	\caption{Sidechain transactions commitment tree. The root hash $h_1$  commits to all sidechain related transactions (for all sidechains) included in the MC block. All $SC\textit{\textbf{X}}Hash$, where $X$ is a sidechain identifier, are ordered by the id and commit to all transactions related to the sidechain $X$. $WCertHash$ commits to the WCert for the sidechain $X$ (if present); only one WCert is allowed for each sidechain. $TxsHash$ commits to FTs and BTRs.}
	\label{fig:4_2}
\end{figure}

To facilitate efficient implementation of the synchronization between the mainchain and sidechains, we modify the structure of a mainchain block header to include an additional value that commits to all sidechain-related actions in the MC block (except the CSW because it is used only when the SC is ceased). This value is a root hash of a Merkle tree that contains all transactions or outputs related to any sidechain (see Fig. \ref{fig:4_2}). We call it the \textit{Sidechain Transactions Commitment} (\textbf{SCTxsCommitment}).

Having $SCTxsCommitment$ in the MC block header allows SC nodes to synchronize and verify SC-related transactions without the need to transmit the entire MC block. Also, it allows the construction of a SNARK proving that all SC-related transactions of the specific MC block have been processed correctly.

\subsection{Bootstrapping Sidechains}

We assume that the mainchain implements a special transaction that allows one to create a sidechain. Such a transaction can be submitted by anyone, and it registers the SC in the mainchain and sets its unique identifier and some system parameters. Once the sidechain is created, a schedule of withdrawal epochs is defined deterministically, and forward/backward transfers must be processed in the mainchain. 

The following set of SC parameters are set upon creation:

\begin{protocolframe}{\textbf{Sidechain configuration}}
\begin{conditions}
    \hspace*{1cm} ledgerId & a unique identifier of the sidechain that has not been used;\\
    \hspace*{1cm} start\_block &  the block number in the mainchain, from which the first withdrawal epoch begins; this parameter defines when the sidechain becomes active;\\
    \hspace*{1cm} epoch\_len & the length of a withdrawal epoch (in MC blocks);\\
    \hspace*{1cm} submit\_len & the period length -- starting from the first block of the withdrawal epoch -- when a withdrawal certificate for the previous epoch must be submitted to the mainchain;\\
    \hspace*{1cm} wcert\_vk & a SNARK verification key $vk_{WCert}$ for WCert proofs;\\
    \hspace*{1cm} btr\_vk & a SNARK verification key $vk_{BTR}$ for BTR proofs;\\
    \hspace*{1cm} csw\_vk & a SNARK verification key $vk_{CSW}$ for CSW proofs;\\
    \hspace*{1cm} wcert\_proofdata & the definition of the \textit{proofdata} structure for the withdrawal certificate; it defines the number and types of included data elements;\\
    \hspace*{1cm} btr\_proofdata & the definition of the \textit{proofdata} structure for the BTR;\\
    \hspace*{1cm} csw\_proofdata & the definition of the \textit{proofdata} structure for the CSW.
\end{conditions}
\end{protocolframe}

Customizable parameters give flexibility in choosing those, which are suitable for a particular sidechain. The triplet \textit{\textbf{(cer\_vk, btr\_vk, csw\_vk)}} is especially important as it defines how the mainchain verifies backward communication from the sidechain. These keys define SNARKs for corresponding operations eventually enabling different designs for sidechains.
\section{The Latus Sidechain} \label{sec:5_Latus_Sidechain}

In the previous section, we described the general sidechain design. Mostly, it was about defining the cross-chain transfer protocol, which provides a communication interface with the mainchain. In this section, we focus on a specific sidechain construction. We give an example of how a decentralized verifiable sidechain can be built on top of the given CCTP.

The general idea is to utilize a recursive composition of SNARKs to construct a succinct proof of the sidechain state progression for the period of a withdrawal epoch. Then, a SNARK for a withdrawal certificate is constructed so that it proves correct sidechain state transition for the whole epoch and validates backward transfers. This allows the mainchain to efficiently verify the sidechain without having to rely on any intermediary -- such as certifiers \cite{GV18} -- and still be oblivious to the sidechain construction and interactions within.

In this section, we provide details of the proposed Latus sidechain that implements decentralized permissionless blockchain $\mathcal{B_{SC}}$ with a proof-of-stake based consensus protocol. We consider $\mathcal{B_{SC}}$  as a simple ledger of payment transactions. We assume that $\mathcal{B_{SC}}$  does not possess its own native asset and, instead, uses only \textit{Coin} asset transferred from the mainchain by means of the CCTP. Additionally, we assume that the mainchain is a classical proof-of-work based blockchain system with Nakamoto consensus \cite{N08} (e.g., Horizen \cite{H19}).

\subsection{Consensus Protocol} \label{sec:5_1_Consensus}

We use a similar consensus protocol as in our previous proposal \cite{GV18} with some minor adjustments. It is based on a modified version of the Ouroboros proof-of-stake consensus protocol~\cite{KRDO17}.

\begin{figure}[htbp]
	\centering
	\includegraphics[trim={1cm 10.55cm 3cm 4.5cm}, clip,width=1\columnwidth] {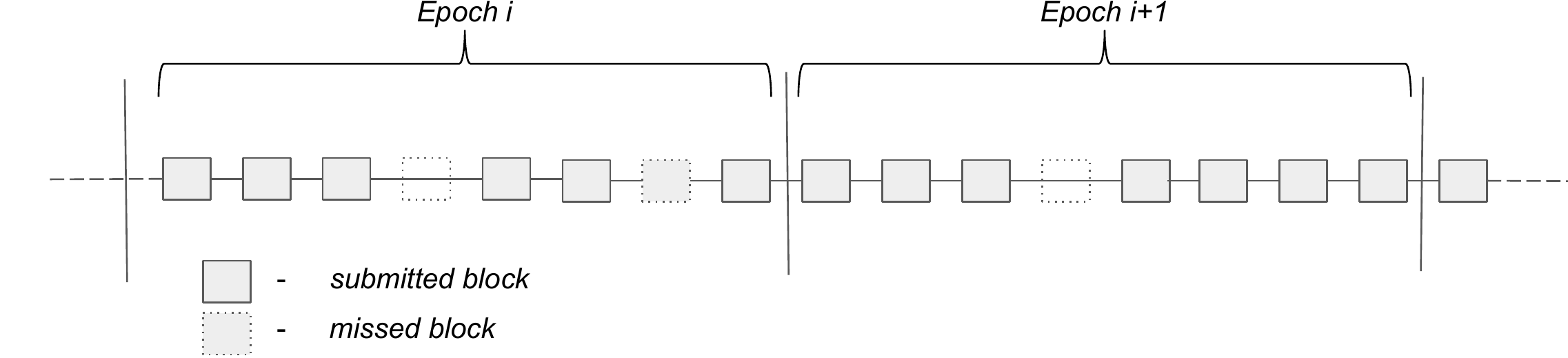}
	\caption{A general scheme of an epoch. Note that even though there is an assigned slot leader for each slot, the leader may skip block generation, and in this case, the slot remains empty.}
	\label{fig:5_1}
\end{figure}

In Ouroboros, time is divided into epochs with a predefined number of slots. Each slot is assigned with a slot leader who is authorized to generate a block during this slot. Slot leaders of a particular epoch are chosen randomly before the epoch begins from the set of all sidechain stakeholders (Fig. \ref{fig:5_1}). The protocol operates in a synchronous environment where each slot takes a specific amount of time (e.g., 20 seconds).

\textbf{Epoch}. An epoch is a sequence of the $k$ successive slots $Ep_i=(sl_i^0,sl_i^1,...,sl_i^{k-1})$, where $k$ is the predefined length of the epoch and $i$ is the epoch sequence number.

\textbf{Slot}. A slot is a specific period in time during which a slot leader is authorized to issue a block. Each slot has the corresponding slot leader who is chosen randomly before the epoch begins. A slot leader may skip generating a block, in this case, the following block will refer to the latest generated block.

\textbf{Slot Leader}. The slot leader of the slot $sl_i^j$ is a stakeholder who was authorized by the Slot Leader Selection Procedure to forge a block at slot $sl_i^j$.

\textbf{Slot Leader Selection Procedure}. The slot leader selection procedure $Select(SD_{Ep_i}, rand)$ is a procedure that selects all slot leaders of the epoch $Ep_i$ according to the fixed stake distribution $SD_{Ep_i}$ and some random value $rand$. The stake distribution $SD_{Ep_i}$ is fixed before the epoch $Ep_i$ begins. The randomness $rand$ is revealed only after the stake distribution is fixed.
\\~\\
In our construction, we additionally introduce binding with the mainchain. This implies that sidechain blocks contain references to mainchain blocks so that their history is preserved in the sidechain. The chain resolution algorithm is altered to enforce that the sidechain follows the longest mainchain branch. 

As a “mainchain block reference”, we consider a whole mainchain block header together with transactions related to the referencing sidechain.

Sidechain block forgers are obliged to keep mainchain references consistent and ordered when included in SC blocks. A sidechain block $SB_j$ can contain a reference to the mainchain block $B_i$ if and only if 

\begin{figure}[htbp]
	\centering
	\includegraphics[trim={1cm 10.57cm 4cm 6.5cm}, clip,width=1\columnwidth] {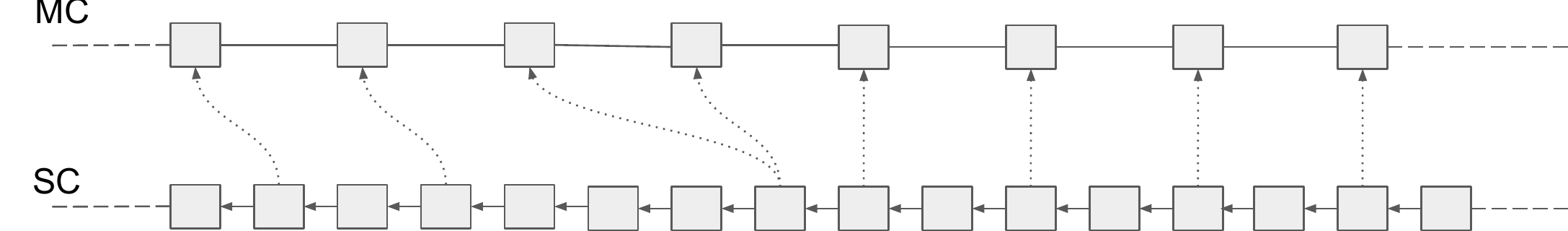}
	\caption{An example of the sidechain binding to the mainchain.}
	\label{fig:5_2}
\end{figure}

\begin{enumerate}[leftmargin=3em, itemsep=0em]
    \item the block $B_i$ is a valid mainchain block, and
    \item references to all previous mainchain blocks $B_k$, $k\in\{\eta,\eta+1,...,i-1\}$ have been already included in sidechain blocks (also considering the current one, as a sidechain block may contain more than one reference), where $\eta$ is the genesis reference (Fig. \ref{fig:5_2}).
\end{enumerate}

Even though it is not mandatory for the block forgers to include mainchain references, we assume that honest block forgers will do this to support the cross-chain transfer protocol between chains. It is also possible to construct an incentive mechanism for block forgers who include references. For instance, users who initiate forward/backward transfers may pay some fee from each transaction. The incentive mechanism is beyond the scope of the current paper as we only provide an example of a sidechain consensus protocol.

The binding to the mainchain provides two important properties of our sidechain construction:

\begin{enumerate}[itemsep=0em]
    \item \textbf{Deterministic synchronization between the MC and the SC}. When the sidechain block $SB_i$ refers to the mainchain block $B_j$, it explicitly acknowledges all transactions included in the block $B_j$. It means that if $B_j$ contains any transactions related to this sidechain (by transactions, we mean forward transfers and backward transfer requests), such transactions are immediately included in the sidechain (see Fig. \ref{fig:5_3}).
    \begin{figure}[htbp]
    \centering
    \begin{minipage}[b]{0.85\textwidth}
        \centering
    	\includegraphics[trim={5cm 11.3cm 5cm 1.65cm}, clip,width=0.9\columnwidth] {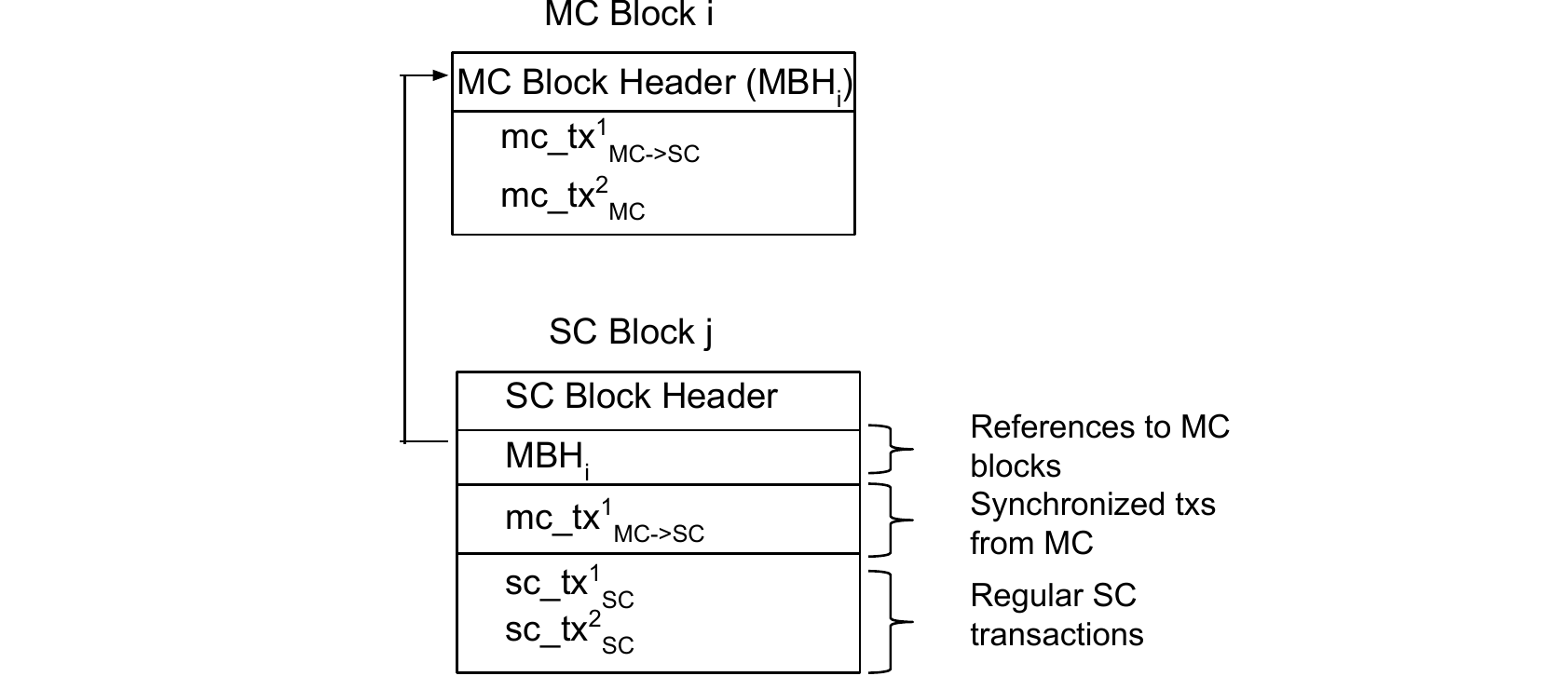}
    	\caption{An example of transaction synchronization between the mainchain and the sidechain: MC block $B_i$ contains one SC-related transaction $mc\_tx_{MC \rightarrow SC}^1$, which is also included in the SC block $SB_j$ because it refers to $B_i$.}
    	\label{fig:5_3}
    \end{minipage}\hfill
    \end{figure}
    
    \item \textbf{Mainchain forks resolution}. It is known that Nakamoto consensus does not provide finality on a chain of blocks \cite{SJS18}. It means that there is always a non-zero probability that some sub-chain of MC blocks will be reverted and substituted by another sub-chain with the more cumulative work. Such behaviour is normally handled by the mainchain but may be disastrous for the sidechain because $MC \rightarrow SC$ transactions that are already confirmed in the sidechain may be reverted in the mainchain. The binding eliminates such situations because in the case of a fork in the MC, SC blocks that refer to forked blocks in the MC would also be reverted.
\end{enumerate}

\textbf{Security}. The standard procedure for proving blockchain consensus protocol security requires demonstrating the ability of the protocol to satisfy two fundamental properties of a distributed ledger: \textbf{liveness} and \textbf{persistence} \cite{GKL14}. Liveness ensures that transactions broadcasted by honest parties will be eventually included in the ledger, and persistence ensures that once a transaction is confirmed by one honest node, it will also be confirmed by all other honest nodes (so that eventually it becomes final and immutable). Such properties are usually proven under certain assumptions, such as honest majority among protocol participants, etc. We refer the interested readers to the original Ouroboros paper \cite{KRDO17} for an exhaustive list of assumptions and properties analysis.

Since the proposed consensus protocol also incorporates binding with the mainchain, it implies an additional assumption of the honest hashing power majority in the mainchain.

We suppose that under these assumptions the proposed protocol derives security guarantees provided by original Ouroboros and Nakamoto consensus protocols.
\\~\\
We want to emphasize that different types of sidechains may adopt different consensus protocols that better suit specific use cases (e.g., fast coin transferring support). A sidechain consensus protocol (including the one described in this section) is not the focus of this research and needs further analysis.

\subsubsection{Withdrawal Epochs}

As it has been described in section [\ref{sec:4_1_2_BackwardTransfers} \nameref{sec:4_1_2_BackwardTransfers}], the Cross-Chain Transfer Protocol introduces the notion of a \textbf{withdrawal epoch}\footnote{Note that withdrawal epochs are independent from epochs in the Ouroboros consensus protocol.} (WE), which is defined as a fixed-length range of MC blocks (length is set upon SC creation). The concept of withdrawal epochs is essential for commanding backward transfers.

\begin{figure}[htbp]
	\centering
	\includegraphics[trim={1cm 9cm 4cm 5.42cm}, clip,width=1\columnwidth] {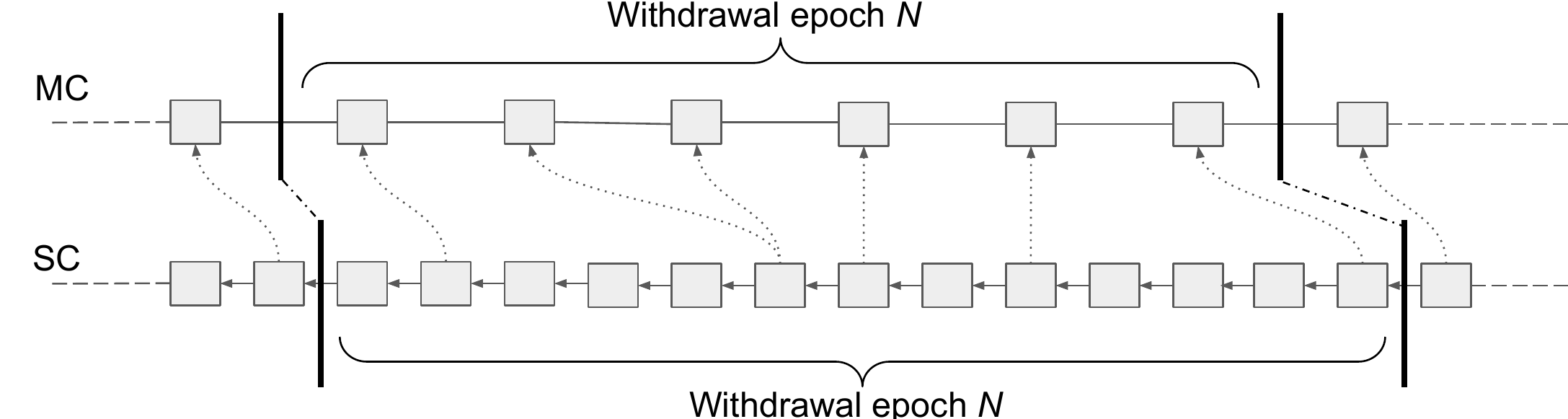}
	\caption{An example of a withdrawal epoch in the sidechain.}
	\label{fig:5_4}
\end{figure}

Following this design, we also introduce withdrawal epochs in a sidechain which coincide with the mainchain withdrawal epochs. A WE is defined as a range of SC blocks where the first and last blocks of the range are determined by references to the first and last MC blocks in the corresponding withdrawal epoch in the MC (see Fig. \ref{fig:5_4}).

Even though a withdrawal epoch in the SC may have variable length (as it depends on when corresponding MC blocks will be referenced), the binding between chains allows to deterministically define the boundaries of the WE in the sidechain.

More formally, if the withdrawal epoch $WE_i^{MC}$ of size $len$ in the MC is defined by a sequence of blocks $WE_i^{MC} = (B_i^0,B_i^1,...,B_i^{len-1})$, then the corresponding withdrawal epoch in the SC can be determined as:
\[WE_i^{SC}=(SB_i^0,SB_i^1,...,SB_i^k),\]
where:

-- $SB_i^0$ is an immediate descendant of the block $SB_{i-1}^n$ which refers to the MC block $B_{i-1}^{len-1}$ (the last one in the withdrawal epoch $WE_{i-1}^{MC}$); and

-- $SB_i^k$ is the block that refers to $B_i^{len-1}$.

Note that to simplify implementation, it might be needed to restrict SC blocks to not refer to several MC blocks on the boundaries of the withdrawal epoch (i.e., if the SC block refers to $B_i^{len-1}$ it cannot also refer to the next MC block $B_{i+1}^0$).

It is important to restate that the notion of the withdrawal epoch is independent from epochs in the Ouroboros consensus protocol. 

\subsection{Accounting Model and System State} \label{sec:5_2_Accounting}

\begin{figure}[htbp]
	\centering
	\includegraphics[trim={2cm 12.6cm 7cm 2cm}, clip,width=0.9\columnwidth] {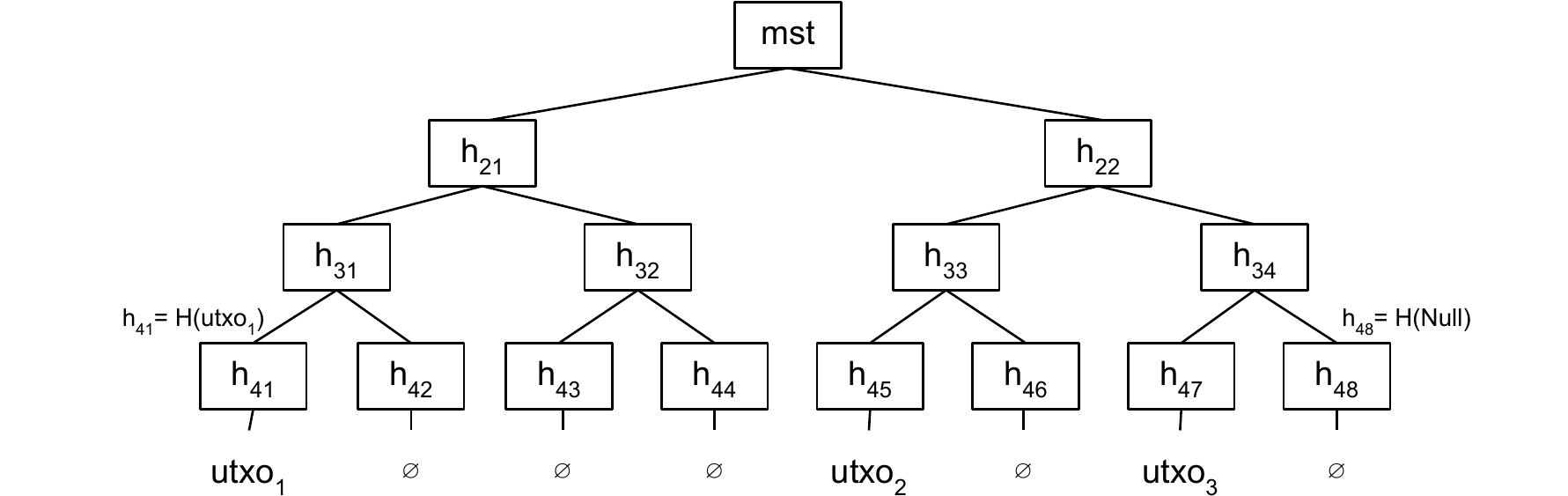}
	\caption{An example of the Merkle State Tree with $D_{MST}=3$. A leaf of the tree is either an unspent output or $Null$ value. The tree contains 3 occupied and 5 empty slots. The function $MST\_Position$ returns the position of a given $utxo$ in the tree, e.g., $MST\_Position(utxo_2)=4$.}
	\label{fig:5_5}
\end{figure}

The Latus blockchain adopts the UTXO-based accounting model \cite{NBFMG16} where the state is represented by a set of unspent outputs combined into a fixed-size Merkle tree (see Fig. \ref{fig:5_5}). We call such a tree a \textbf{Merkle State Tree (MST)}. Lowercase $mst_t$ stands to denote the root hash of the $MST_t$  tree at the moment $t$. 

The depth $D_{MST}$ of the MST tree is a fixed system parameter that also constrains the total number of UTXOs that can exist in the system to be at most $2^{D_{MST}}$ (see Fig. \ref{fig:5_5}).

We consider each leaf of the MST as a UTXO slot that can be “\textit{occupied}” or “\textit{empty}” at a given moment. We introduce the deterministic function $MST\_Position(utxo_i)$ that returns the position of some unspent output $utxo_i$ if it is included in the tree. Note that the $utxo$ position does not depend on the current state of the MST.

The unspent transaction output (UTXO) is defined as a tuple $(addr,amount,nonce)$ where 
\begin{itemize}[leftmargin=4em, itemsep=-0.3em]
    \item $addr$ is an address of the UTXO owner who possesses the corresponding private key that allows to spent it;
    \item $amount$ is the number of coins secured by the UTXO; and
    \item $nonce$ is a unique identifier of the UTXO.
\end{itemize}

\subsubsection{System State}

Provided with the MST structure, which is the core of a sidechain state, we define an overall SC system state at the moment $t$ as a tuple:
\[state_t\ \stackrel{\mathrm{def}}{=}\ (MST_t,\ backward\_transfers_t),\]
where $backward\_transfers$ is a list of backward transfers initiated in the current withdrawal epoch. $backward\_transfers$ is transient and reset every new withdrawal epoch.
\subsection{Transactional Model} \label{sec:5_3_TransactionalModel}

There are 4 types of transactions defined in the Latus sidechain that realize basic payment functionality and cross-chain transfer protocol. To simplify the model, we consider them as logical transactions, though we stress that a real-world implementation can be optimized so that a single transaction on the blockchain may combine several logical transactions (even with different types). 

The transactions are the following:
\begin{enumerate}[leftmargin=4em,itemsep=0em]
    \item \textbf{Payment (PTx)} -- transfers coins within the sidechain.
    \item \textbf{Backward Transfer (BTTx)} -- initiates transfer of coins from the SC to the MC.
    \item \textbf{Forward Transfers (FTTx)} - receives coins transferred from the mainchain.
    \item \textbf{Backward Transfer Requests (BTRTx)} - initiates coin transfer from the SC to the MC. In contrast to \textit{BTTx}, \textit{BTRTx} contains BTRs initially submitted in the mainchain and then synchronized to the SC.
\end{enumerate}

Whereas \textit{PTx} and \textit{BTTx} are inherently SC-defined transactions (thus, submitted and processed in the sidechain), \textit{FTTx} and \textit{BTRTx} are MC-defined transactions (they encapsulate FTs and BTRs that are initially submitted to the MC). We describe each type in detail in the following sections.

\subsubsection{Payment Transaction}

We define a regular payment as a multi-input multi-output transaction \cite{BitWiki-Tx}:
\begin{protocolframe}{}
\begin{lstlisting}
type PaymentTx {
	inputs: List[UTXO];
	signatures: List[Signature];
	outputs: List[UTXO];
}
\end{lstlisting}
\end{protocolframe}
where:
\begin{enumerate}[itemsep=-0.1em]
    \item \textit{inputs} are some unspent outputs from previous transactions, spending of which are authorized by \textit{signatures}, and
    \item the total coin’s value of \textit{inputs} is equal or greater than the total coin’s value of \textit{outputs}.
\end{enumerate}

The state transition function $update$ for the payment transaction is defined in the following way:
\[state_{i+1}=update(tx_{pay},state_i),\]
where $state_{i+1}[backward\_transfers]$ is unchanged and $state_{i+1}[MST]$ is derived from $state_i[MST]$ by
\begin{enumerate}[itemsep=-0.1em]
    \item removing all UTXOs that are inputs in $tx_{pay}$ and substituting them with $Null$ to produce $MST_i^-$; and
    \item sequentially adding to $MST_i^-$ all UTXOs that are outputs in $tx_{pay}$ according to \\ $MST\_Position(utxo_j)$.
\end{enumerate}

\subsubsection{Forward Transfers Transaction} \label{sec:5_3_2_FTTx}

Forward transfers allow one to send coins from the mainchain to a sidechain. As such, FTs are first submitted to the MC and processed there (destroying coins) and then, by means of deterministic synchronization, are included and processed in the sidechain. Recall from [\ref{sec:4_1_1_ForwardTransfers} \nameref{sec:4_1_1_ForwardTransfers}] the basic structure of a forward transfer on the mainchain side:
\[FT\ \stackrel{\mathrm{def}}{=}\ (ledgerId,\ receiverMetadata,\ amount).\]

$receiverMetadata$ is defined by the sidechain construction and in Latus it is just a receiver address and a payback address on the MC needed in case of transfer failure: 

\[receiversMetadata\ \stackrel{\mathrm{def}}{=}\ (receiverAddr,\ paybackAddr).\]

A single MC block may contain several forward transfers related to different sidechains. The sidechain will synchronize FTs present in the referenced MC block by including a special $ForwardTransfers$ transaction (FTTx) in the SC block. Such FTTx specifies all forward transfers from the referenced MC block that are related to this specific sidechain. From the sidechain perspective, we can consider FTTx as a coinbase transaction (the one that creates new coins \cite{BitWiki-Cb}) that is authorized by the mainchain.

We assume that a particular forward transfer may fail so that coins cannot be received by the sidechain. In this case, coins are sent back to the mainchain by creating a corresponding backward transfer. It is done automatically upon FTTx execution in the sidechain.

The reasons for FT failure can be different. For instance, FT's $receiverMetadata$ may be malformed (recall that the MC does not validate semantics of FT’s $receiverMetadata$) or some other sidechain-specific failures occur (e.g., it may happen that $MST\_Position(output_{new})$ maps newly created output to an already occupied slot in $MST_i$, thus causing a collision).

The basic structure of the \textit{ForwardTransfers} transaction is the following:
\begin{protocolframe}{}
\begin{lstlisting}
type ForwardTransfersTx (mcid: BlockID, ft: List[FT]) {
	outputs: List[UTXO];
	rejectedTransfers: List[BackwardTransfer]
}
\end{lstlisting}
\end{protocolframe}
\noindent
where:
\begin{conditions}
    mcid & an identifier of the MC block whose forward transfers are synchronized;\\
    ft & a list of forward transfers from the MC block $mcid$ related to the sidechain where FTTx occurs;\\
    outputs & outputs created for the transferred coins; each valid forward transfer spawns a corresponding output with the same amount of coins;\\
	rejectedTransfers & a list of backward transfers for failed forward transfers; each failed forward transfer spawns a corresponding backward transfer with the same amount of coins.
\end{conditions}

The state transition function update for the FTTx is defined in the following way:
\[state_{i+1}=update(tx_{FT},state_i),\]
\vbox{
\noindent
where:
\begin{enumerate}[itemsep=-0.1em]
    \item $state_{i+1}[MST]$ is derived from $state_i[MST]$ by sequentially adding all UTXOs that are outputs in $tx_{FT}$ according to $MST\_Position(utxo_j)$;
    \item $state_{i+1}[backward\_transfers]$ is derived from $state_i[backward\_transfers]$ by appending $rejectedTransfers$ from $tx_{FT}$.
\end{enumerate}}

Note that failed forward transfers are recovered with the backward transfer mechanism through a withdrawal certificate at the end of the epoch. Recall that the MC knows nothing about the SC state and cannot know that an FT is failed; thus, we use the standard mechanism to reclaim coins in the MC.

\subsubsection{Backward Transfer Transaction} \label{sec:5_3_3_BTTx}

A backward transfer transaction (BTTx) allows one to create a request for a backward transfer in the sidechain that will be included in the next withdrawal certificate and then passed and processed in the mainchain.
\begin{protocolframe}{}
\begin{lstlisting}
type BackwardTransferTx {
	inputs: List[UTXO];
	signatures: List[Signature];
	backwardTransfers: List[BackwardTransfer];
}
\end{lstlisting}
\end{protocolframe}
\noindent
where:
\begin{enumerate}[itemsep=-0.1em]
    \item \textit{inputs} are some unspent outputs from previous transactions (spending of which is authorized by \textit{signatures});
    \item \textit{backwardTransfers} are data about receivers of coins on the mainchain side; recall from [\ref{sec:4_1_2_BackwardTransfers} \nameref{sec:4_1_2_BackwardTransfers}] that the basic structure of a backward transfer is \\$BT\ \stackrel{\mathrm{def}}{=}\ (receiverAddr,\ amount)$;
    \item the total coin value of \textit{inputs} is equal or greater to the total coin value of \textit{backwardTransfers}.
\end{enumerate}

The state transition function $update$ for a backward transfer transaction is defined as:
\[state_{i+1}=update(tx_{BT},state_i),\]
where:
\begin{enumerate}[itemsep=-0.1em]
    \item $state_{i+1}[MST]$ is derived from $state_i[MST]$ by removing all UTXOs that are inputs in $tx_{BT}$; and
    \item $state_{i+1}[backward\_transfers]$ is derived from $state_i[backward\_transfers]$ by appending $backwardTransfers$ from $tx_{BT}$.
\end{enumerate}

Essentially, we can consider $backwardTransfers$ in $tx_{BT}$  as specialized outputs that are unspendable on the sidechain but used to reclaim coins in the mainchain (when transferred by means of a withdrawal certificate). In this respect, BTTx transaction is a special case of regular payment transaction where all outputs are backward transfers.

More details about the entire backward transfer flow can be found in [\ref{sec:5_5_3_BackwardTransfers} \nameref{sec:5_5_3_BackwardTransfers}].

\subsubsection{Backward Transfer Requests Transaction} \label{sec:5_3_4_BTRTx}

The \textit{Backward Transfer Request} (BTR), which is submitted to the MC, is similar to BTTx in the sense that it allows one to create a request that will result in a backward transfer in the next withdrawal certificate if the request is legitimate (e.g., claimed coins were present at the moment of BTR inclusion in the SC block). The difference from BTTx is that BTR is submitted in the mainchain and is used in situations when BTTx cannot be used for some reason. Recall from [\ref{sec:mainchain_managed_withdrawals} \nameref{sec:mainchain_managed_withdrawals}] the basic structure of the BTR on the mainchain side:

\[BTR\ \stackrel{\mathrm{def}}{=}\ (ledgerId,\ receiver,\ amount,\ proofdata,\ proof).\]

\textit{proofdata} and SNARK \textit{proof} are defined by sidechain construction. In Latus, \textit{proofdata} contains an unspent output that should be consumed in the SC to provide coins for transferring: 
\[proofdata\ =\ \{utxo\}.\]

The spending right for the \textit{utxo} should be enforced by the \textit{proof} which is validated upon submission in the MC. 

Similar to forward transfers, a single MC block may contain several BTRs. The sidechain synchronizes BTRs by including in the SC block a special \textit{BackwardTransferRequests} transaction (BTRTx) that contains all BTRs relevant to this sidechain from the referenced MC block. From a sidechain perspective, we can consider BTRTx as an aggregated transaction where each BTR represents a separate backward transfer.

Some BTRs from BTRTx may be invalid when they are synced to the sidechain (e.g., a malicious user may try to spend the same utxo directly in the sidechain before BTR is synced (double-spend problem). Such BTRs are rejected by the sidechain (rejection means that they do not spawn corresponding backward transfers and do not affect the state).

The basic BTRTx structure on the SC side is the following: 
\begin{protocolframe}{}
\begin{lstlisting}
type BackwardTransferRequestsTx (mcid: BlockId, btr: List[BTR]) {
    inputs: List[UTXO];
    backwardTransfers: List[BackwardTransfer];
}
\end{lstlisting}
\end{protocolframe}
\noindent
where:
\begin{conditions}
    mcid & an identifier of the MC block whose BTRs are synchronized;\\
	btr & a list of backward transfer requests from the MC block $mcid$ related to this sidechain;\\
	inputs & a combined list of UTXOs derived from $btr.proofdata$ of each valid BTR;\\
    backwardTransfers & a  list of backward transfers for valid BTRs.
\end{conditions}

The state transition function update for a BTRTx transaction is:
\[state_{i+1}=update(tx_{BTR},state_i),\]
where:
\begin{enumerate}[itemsep=-0.1em]
    \item $state_{i+1}[MST]$ is derived from $state_i[MST]$ by removing all UTXOs that are~inputs~in~$tx_{BTR}$;
    \item $state_{i+1}[backward\_transfers]$ is derived from $state_i[backward\_transfers]$ by appending $backwardTransfers$ from $tx_{BTR}$.
\end{enumerate}

Note that correct processing of BTRs in the sidechain is to be enforced by a withdrawal certificate SNARK proof.
\subsection{State Transition Proof} \label{sec:5_4_StateTransitionProof}

In [\ref{sec:5_3_TransactionalModel} \nameref{sec:5_3_TransactionalModel}], we defined four types of transactions that represent basic state transitions in our sidechain system. Given that all transactions are applied sequentially in an order defined by blocks containing them, we can consider a merged state transition for a sequence of transactions from several blocks:
\[state_{i+k}=update([tx_1,...,tx_k], state_i)=update(tx_k, update(tx_{k-1}, ..... update(tx_1,state_i)).\]

In particular, we are interested in merging transitions for the whole withdrawal epoch and proving that the top-level merged transition is correct. It can be accomplished using the recursive SNARKs composition for state transitions defined in [Def. \ref{def:2_5_rec_snarks}].

The main idea is to construct a single SNARK proof of transition for the whole withdrawal epoch which then can be attached to a withdrawal certificate proving to the mainchain the validity of everything that has happened in the sidechain -- including certificate backward transfers -- without actually revealing any details except state snapshots (in a form of simple hashes) before and after transition.  

We do not go deeply into the details of the SNARKs architecture which is quite sophisticated in this case and requires separate writing to be properly explained; instead, we are going to provide the basic idea of constructing such proof and how it is going to be used.

Let us denote by $s_i=H(state_i)$ the hash value that represents $state_i$. Note that it must be an efficient hashing procedure as it should be implemented for a SNARK arithmetic constraint system. For instance, we can consider $H(\cdot)$ as a root hash of a Merkle tree that contains all the data from $state_i$.

Let us assume that for each basic state transition (represented by $tx_{pay}$, $tx_{FT}$, $tx_{BT}$, and $tx_{BTR}$) we have a corresponding \textit{Base} SNARK [Def. \ref{def:2_5_rec_snarks}] which proves the correct state transition for a single $tx_a,\  a\in{\{pay, FT, BT, BTR\}}$:

\[\pi_a^{Base} \leftarrow Prove(pk_a^{Base}, (s_i,s_{i+1}), (tx_a)),\]
\vspace*{-6mm}
\[true/false \leftarrow Verify(vk_a^{Base}, (s_i,s_{i+1}), \pi_a^{Base}).\]

Also, let us assume that we have a \textit{Merge} SNARK which takes two proofs of adjacent state transitions (\textit{Base} or \textit{Merge}) and combines them into a single proof:

\[\pi^{Merge} \leftarrow Prove(pk^{Merge}, (s_i,s_{i+k}), (s_{i+j},\pi_1^b,\pi_2^c)),\]
\vspace*{-6mm}
\[true/false \leftarrow Verify(vk^{Merge}, (s_i,s_{i+k}), \pi^{Merge}).\]
where:
\begin{itemize}[leftmargin=3em,itemsep=0em]
    \item $b,c\in{\{Base,Merge\}}$;
    \item $\pi_1^b$ proves that there exist such $tx_1,...,tx_j$ so that $state_{i+j}=update([tx_1,...,tx_j], state_i)$;
    \item $\pi_2^c$ proves that there exist such $tx_{j+1},...,tx_k$ so that $state_{i+k}=update([tx_{j+1},...,tx_k], state_{i+j})$.
\end{itemize}

\begin{figure}[htbp]
	\centering
	\includegraphics[trim={2cm 6.4cm 2cm 3cm}, clip,width=1\columnwidth] {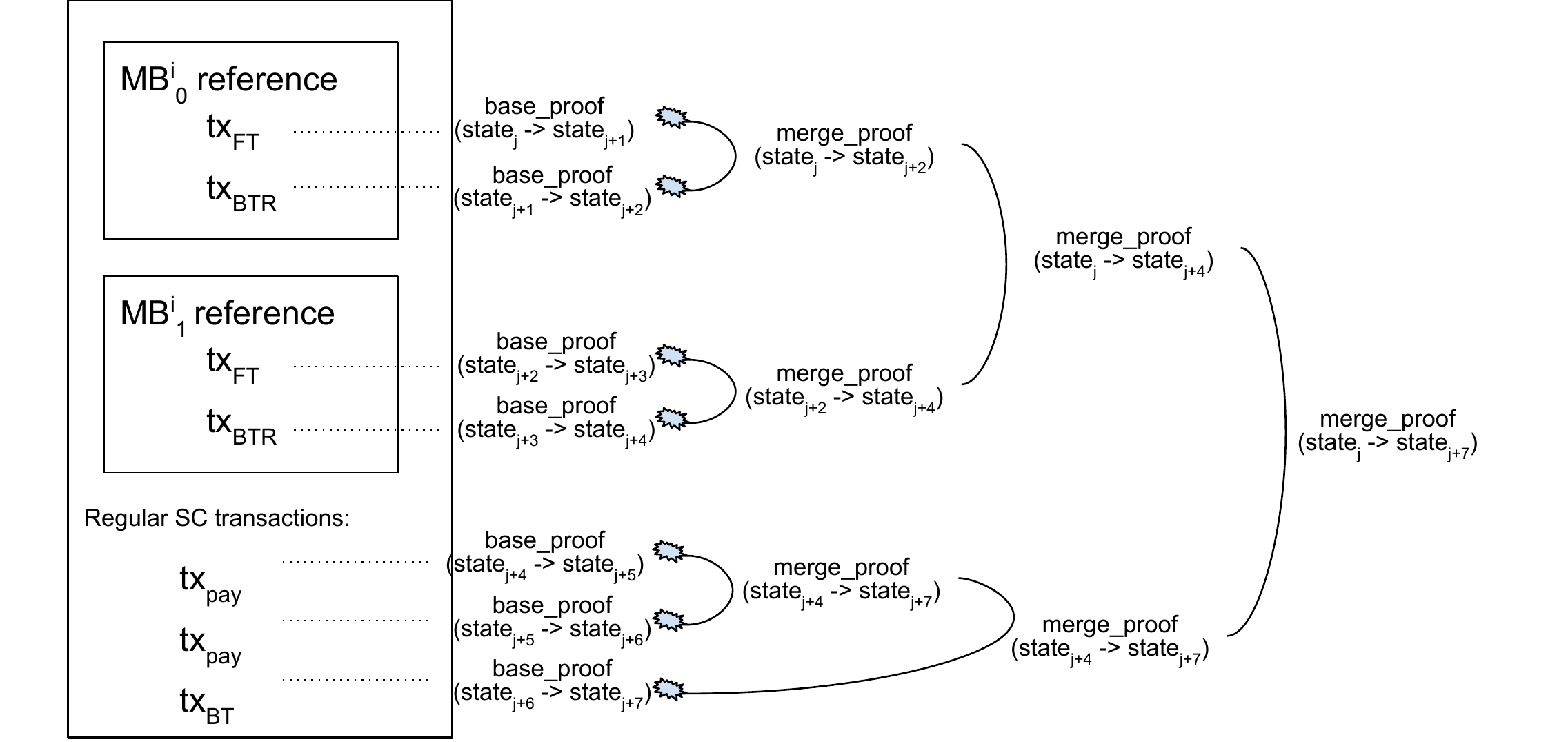}
	\caption{Recursive composition of state transition proofs for the whole SC block. At the bottom level, there are proofs for basic transitions (represented by transactions included in the block) which are then recursively merged into a single proof.}
	\label{fig:5_6}
\end{figure}

\begin{figure}[htbp]
	\centering
	\includegraphics[trim={0cm 9cm 0.5cm 3.5cm}, clip,width=1\columnwidth] {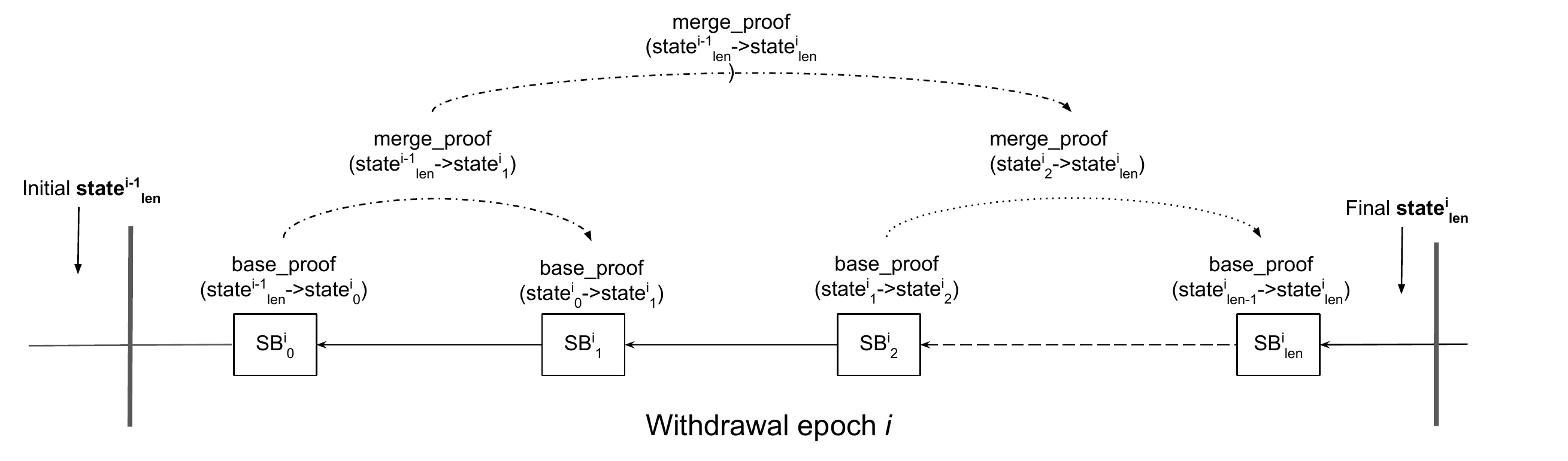}
	\caption{Recursive composition of state transition proofs for the whole withdrawal epoch. State transitions for SC blocks are considered as base transitions though they themselves are recursively constructed from basic transitions (see Fig. \ref{fig:5_6}).}
	\label{fig:5_7}
\end{figure}

Provided with this construction, we can recursively build a single SNARK proof of state transition for a whole withdrawal epoch from the sequence of basic transitions. This process is visualized in figures \ref{fig:5_6} and \ref{fig:5_7}.

Figure \ref{fig:5_6} demonstrates the recursive construction of a state transition proof for a single sidechain block. Note that the scheme is simplified; in reality, the SNARKs composition is more sophisticated and the proof itself attests not only for the correctness of basic transitions but also for the validity of the SC block, the validity of included MC block references, their contiguity, etc.

Figure \ref{fig:5_7} demonstrates the recursive construction of a state transition proof for an entire withdrawal epoch. Provided with the proofs of state transitions for blocks from the previous step, now they are merged to generate a single proof for the whole epoch which is used to construct a final proof for a withdrawal certificate. 

In a nutshell, each withdrawal certificate $WCert_i$ for epoch $i$ commits to the new state $state_{len}^i$ produced by applying all blocks belonging to epoch $i$ and proves correct transition from the $state_{len}^{i-1}$ committed by the previous withdrawal certificate. This also involves proving that all MC blocks belonging to the withdrawal epoch are referenced and all MC transactions related to this sidechain are processed. As forward and backward transfers are among basic transitions they will also be proven.

Again, we stress that this description is greatly simplified just to show the basic idea of recursive SNARKs composition for state transitions.

\subsubsection{Performance and Incentives}

Generating a SNARK proof for each basic transition and then merging them together requires a significant amount of computation. This task cannot be solely levied upon forgers or WCert issuers. Currently, we are investigating different approaches. 

One of the possible solutions is to introduce a special dispatching scheme that assigns generation of proofs randomly to interested parties who then do these tasks in parallel and submit generated proofs to the blockchain. An incentive scheme provides a reward for each valid submission.

We consider this as a separate topic of research which we do not elaborate in this paper.

\subsection{Cross-Chain Transfer Protocol}

In the previous sections, we described the consensus protocol, accounting model, and transactional model of the proposed sidechain construction. In this section, we will focus on the structure of the cross-chain transfer protocol on the sidechain side which is based on those components.

\subsubsection{Mainchain Block Reference} \label{sec:5_5_1_MCBlockRef}

In [\ref{sec:5_1_Consensus} \nameref{sec:5_1_Consensus}], we briefly described the synchronization procedure between the mainchain and sidechain, which relies on MC block referencing. Here, we describe the reference structure in a more detailed way.

Recall that an MC block header contains the \textbf{SCTxsCommitment} [\ref{sec:4_1_3_ScTxsCommitment} \nameref{sec:4_1_3_ScTxsCommitment}] field that commits to all SC-related transactions/outputs in that block:

\begin{protocolframe}{}
\begin{lstlisting}
type MCBlockHeader {
    prevBlock: BlockId
    height: Int
    ...
    scTxsCommitment: Hash
    ...
}
\end{lstlisting}
\end{protocolframe}

\begin{figure}[htbp]
	\centering
	\includegraphics[trim={0.5cm 8.7cm 8cm 1.6cm}, clip,width=0.9\columnwidth] {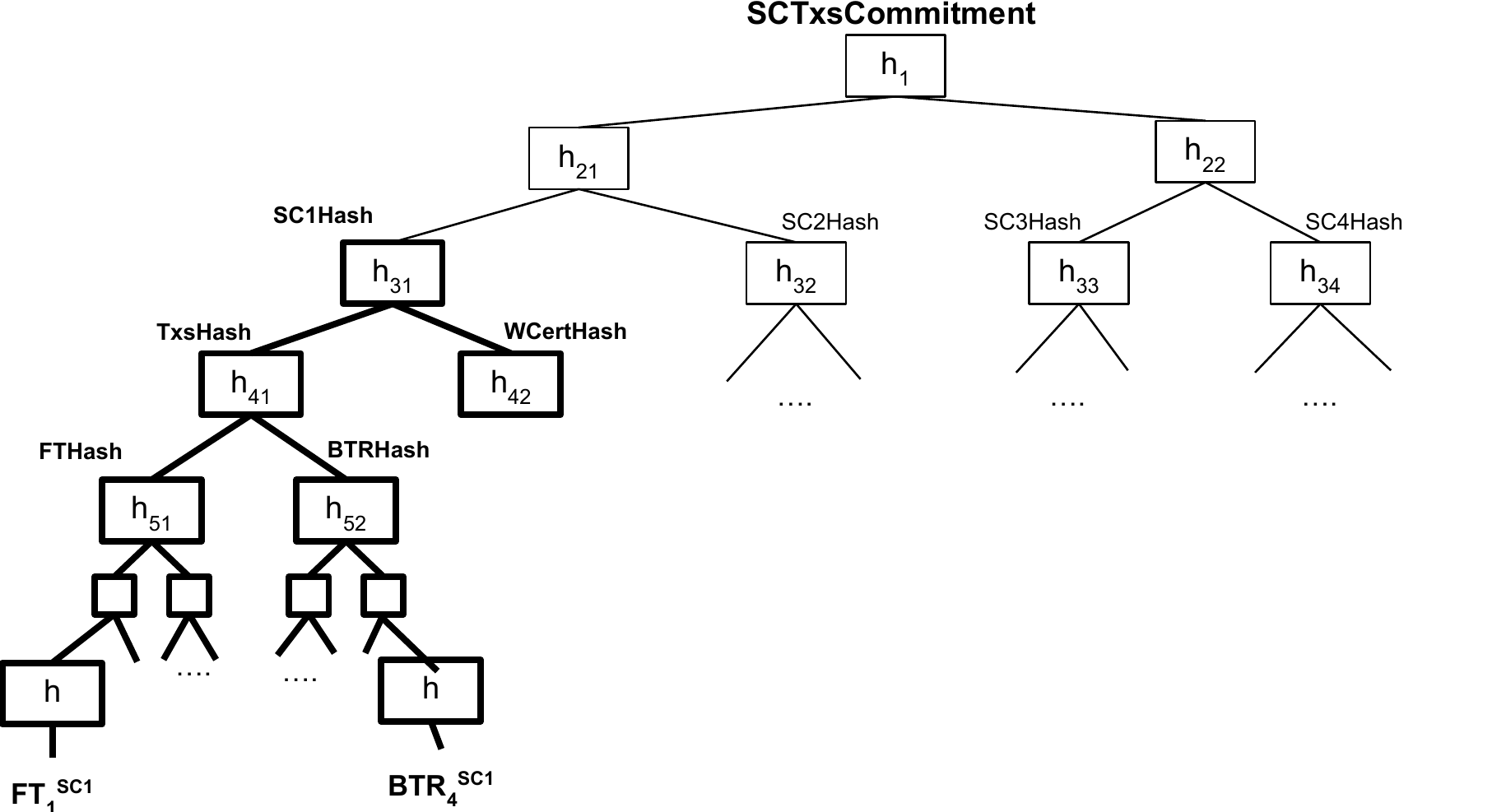}
	\caption{An example of the sidechain transactions commitment tree. One of the intermediate nodes (\textbf{SC1Hash}) is a root hash of the subtree that commits to all transactions related to the sidechain $SC1$.}
	\label{fig:5_8}
\end{figure}

ScTxsCommitment is a root hash of a Merkle tree where one of the subtrees is the Merkle tree of transactions related to the sidechain that referenced the block (see Fig. \ref{fig:5_8}).

The structure of the mainchain block reference is the following:
\begin{protocolframe}{}
\begin{lstlisting}
type MCBlockReference {
    header: MCBlockHeader
    mproof: Option[MerkleProof]
    proofOfNoData: Option[MerkleProof[]] 
    forwardTransfers: Option[FTTx]
    btRequests: Option[BTRTx]
    wcert: Option[WCert]
}
\end{lstlisting}
\end{protocolframe}
\vbox{where:
\begin{conditions}
    header & a header of the MC block that is referenced;\\
	
	mproof & optional field: in case the MC block includes at least one transaction related to this SC, $mproof$ will contain a Merkle proof [Def. \ref{def:MHT}] for the intermediate node in the sidechain transactions commitment tree that is a root of a subtree of transactions related to this sidechain (by the example in Fig. \ref{fig:5_8}: the subtree root for the sidechain $SC1$ is $h_{31}$ and the corresponding Merkle proof is the tuple of nodes $\{h_{32},h_{22}\}$); in case the MC block has no transactions related to this SC, the $mproof$ must be $Null$;\\
	
	proofOfNoData & optional field: in case the MC block has no transactions related to this SC, $proofOfNoData$ contains the Merkle proof(s) necessary to prove that this $ledgerId$ was not part of the $SCTxsCommitment$ tree;\\

	forwardTransfers & optional field: it is either a \textit{ForwardTransfers} transaction [\ref{sec:5_3_2_FTTx} \nameref{sec:5_3_2_FTTx}] (if the MC block contains at least one forward transfer to this sidechain) or otherwise $Null$;\\
	
	btRequests & optional field: it is either a \textit{BackwardTransferRequests} transaction [\ref{sec:5_3_4_BTRTx} \nameref{sec:5_3_4_BTRTx}] (in case the MC block contains at least one backward transfer request to this sidechain) or otherwise $Null$;\\
	
	wcert & optional field: it is either a withdrawal certificate (in case the MC block contains the withdrawal certificate related to this sidechain) or otherwise $Null$.
\end{conditions}}

Provided with \textit{mproof}, \textit{forwardTransfers}, \textit{btRequests}, and \textit{wcert} fields, the $SCTxsCommitment$ can be reconstructed and verified against the \textit{scTxsCommitment} field included in the MC block header. It allows to verify that all SC-related transactions were correctly synchronized from the MC block without the need to download and verify its body. Moreover, we can construct a SNARK proving that the MC block reference has been correctly processed and that all SC-related transactions have been applied -- it is an essential part of constructing a state transition proof for a withdrawal epoch [\ref{sec:5_4_StateTransitionProof} \nameref{sec:5_4_StateTransitionProof}].

\subsubsection{Forward Transfers}

In [\ref{sec:4_1_1_ForwardTransfers} \nameref{sec:4_1_1_ForwardTransfers}] and [\ref{sec:5_3_2_FTTx} \nameref{sec:5_3_2_FTTx}], we have already discussed most of the details related to the forward transfer design both on the mainchain and sidechain sides. Here we combine everything.

In general, it looks as follows: an MC to SC transfer is represented by a pair of transactions which we can consider as “\textit{sending}” and “\textit{receiving}”. “\textit{Sending}” is done on the mainchain side by means of the forward transfer defined in [\ref{sec:4_1_1_ForwardTransfers} \nameref{sec:4_1_1_ForwardTransfers}] and “\textit{receiving}” is done on the sidechain side by means of aggregated \textit{ForwardTransfers} transaction defined in [\ref{sec:5_3_2_FTTx} \nameref{sec:5_3_2_FTTx}]. While “\textit{sending}” destroys coins in the mainchain, “\textit{receiving}” creates the corresponding number of coins in the sidechain.

\begin{figure}[htbp]
	\centering
	\includegraphics[trim={1cm 9.63cm 1cm 4.33cm}, clip,width=0.9\columnwidth] {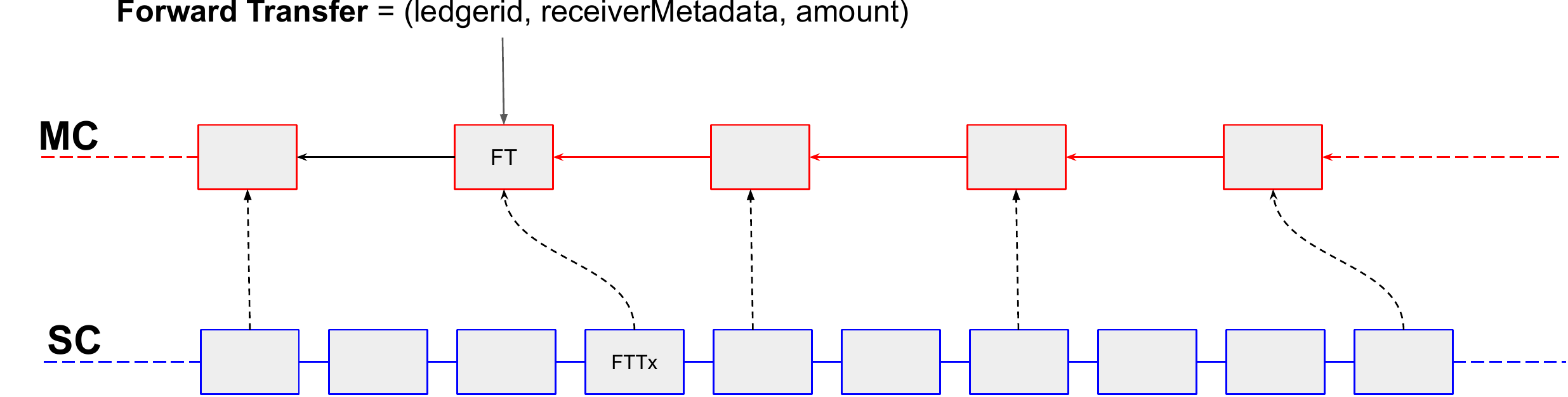}
	\caption{Forward transfers syncing from the mainchain to the sidechain.}
	\label{fig:5_9}
\end{figure}

Forward transfers submitted to the mainchain become available in the sidechain at the moment the MC block containing them is referenced in the sidechain (see Fig. \ref{fig:5_9}). With the MC block reference [\ref{sec:5_5_1_MCBlockRef} \nameref{sec:5_5_1_MCBlockRef}], a \textit{ForwardTransfers} transaction (FTTx) is included in the SC block (if there are any FTs).

The consistency of forward transfers included in a sidechain FTTx is verified by recalculating the \textit{FTHash} (Fig. \ref{fig:5_8}) and checking the \textbf{SCTxsCommitment} following the procedure described in  [\ref{sec:5_5_1_MCBlockRef} \nameref{sec:5_5_1_MCBlockRef}].

\subsubsection{Backward Transfers} \label{sec:5_5_3_BackwardTransfers}

In general, backward transfer is a transfer of coins in the opposite direction: from the sidechain to the mainchain. This operation is more sophisticated and thus requires several sub-protocols to provide sufficient security and reliability.

There are three ways to withdraw coins from the sidechain to the mainchain:
\begin{enumerate}[leftmargin=3em,itemsep=0em]
    \item \textbf{Regular withdrawal} is a standard mechanism that is used under normal conditions. It implies the usage of a backward transfer transaction [\ref{sec:5_3_3_BTTx} \nameref{sec:5_3_3_BTTx}] and a withdrawal certificate to transfer coins to the mainchain.
    \item \textbf{Backward transfer request} is similar to the regular withdrawal with that difference that it is initially submitted to the mainchain [\ref{sec:mainchain_managed_withdrawals} \nameref{sec:mainchain_managed_withdrawals}] and then synchronized to the sidechain by means of [\ref{sec:5_3_4_BTRTx} \nameref{sec:5_3_4_BTRTx}]. The coins are transferred to the mainchain with a withdrawal certificate.
    \item \textbf{Ceased sidechain withdrawal} is a mechanism that is used when the sidechain is no longer operating. This type of withdrawal does not use withdrawal certificates and supposes direct handling by the mainchain.
\end{enumerate}

\begin{figure}[htbp]
	\centering
	\includegraphics[trim={1cm 9.65cm 1cm 4cm}, clip,width=0.9\columnwidth] {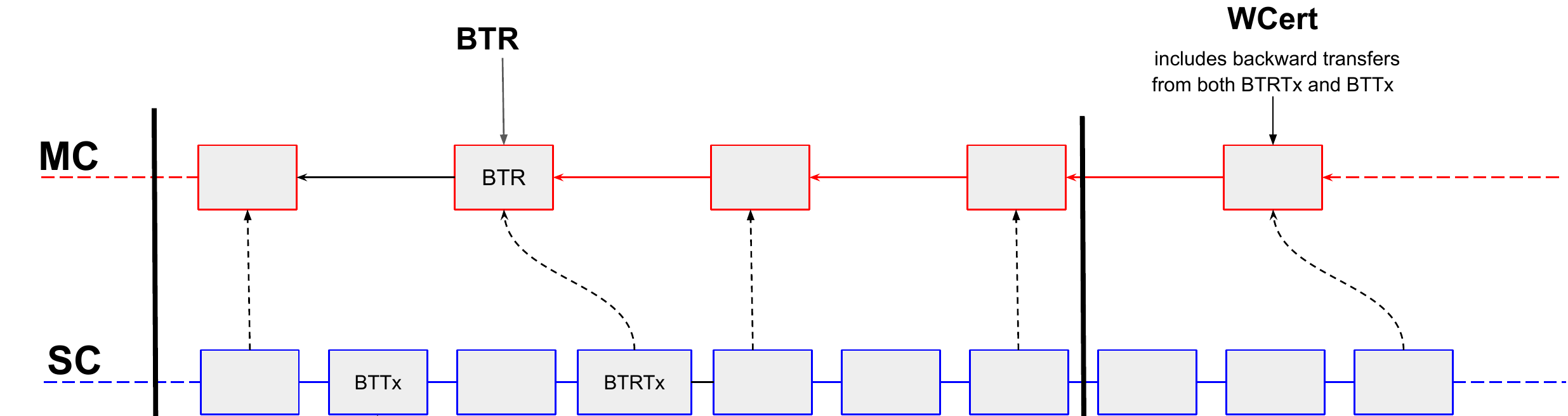}
	\caption{Withdrawing coins with BT and BTR transactions.}
	\label{fig:5_10}
\end{figure}

The first two types of withdrawals (regular and BTR) use the standard mechanism for backward transfers - \textbf{withdrawal certificate}. Most of the details related to their submission and processing have already been discussed in [\ref{sec:mainchain_managed_withdrawals} \nameref{sec:mainchain_managed_withdrawals}], [\ref{sec:5_3_3_BTTx} \nameref{sec:5_3_3_BTTx}], and [\ref{sec:5_3_4_BTRTx} \nameref{sec:5_3_4_BTRTx}]. The basic principle is summarized in figure \ref{fig:5_10}.

\textbf{Regular withdrawal}. A special \textit{BackwardTransfer} transaction is submitted to the sidechain by a user who wants to transfer coins. This transaction destroys coins in the sidechain. At the end of the withdrawal epoch, all backward transfers are collected in a withdrawal certificate which is submitted to the mainchain where it is processed, and the corresponding number of coins is created in the mainchain.

\textbf{Backward transfer request}. BTRs are submitted to the mainchain and synchronized to the sidechain by means of a \textit{BackwardTransferRequests} transaction (analogously to forward transfers). The consistency of BTRs included in the sidechain is verified by recalculating the \textit{BTRHash} (Fig. \ref{fig:5_8}) and checking its presence in the SCTxsCommitment tree following the procedure described in  [\ref{sec:5_5_1_MCBlockRef} \nameref{sec:5_5_1_MCBlockRef}]. After the BTR is synchronized to the sidechain, it is processed as regular withdrawal through a withdrawal certificate. 
\\~\\
In the following sections, we will discuss more deeply the structure and generation of a withdrawal certificate and BTR. We will also separately discuss CSW as it is conceptually different from the first two withdrawal methods.

\subsubsubsection{Withdrawal Certificate} \label{sec:5_5_2_1_WCert}

Withdrawal certificate is a pivotal component of the backward transfer flow. Recall the basic structure of a withdrawal certificate that is defined by the mainchain [\ref{sec:4_1_2_BackwardTransfers} \nameref{sec:4_1_2_BackwardTransfers}]:

\[WCert\ \stackrel{\mathrm{def}}{=}\ (ledgerId,\ epochId,\ quality,\ BTList,\ proofdata,\ proof).\]

While \textit{ledgerId} and \textit{epochId} are global parameters known to the mainchain, the semantics of \textit{quality}, \textit{proofdata}, and \textit{proof} are defined by the sidechain. The withdrawal certificate is created once per a withdrawal epoch and includes all backward transfers that have been submitted during the epoch.
\\~\\
\textbf{Quality}. The quality parameter is used by the mainchain to determine what WCert should be adopted in case several have been submitted for the same epoch. From the MC side, this is just an integer value which can be compared with quantities from other certificates. In the Latus sidechain, we define the quality to be the height of the blockchain up until which the WCert proves state transition.
\\~\\
\textbf{Backward Transfers List}. BTList is a list of backward transfers collected during a withdrawal epoch for which the certificate is created:
\[BTList = state_i[backward\_transfers], \]
where $state_i$ is the state of the sidechain after applying the last block in the withdrawal epoch.

\begin{center}
    \textbf{Withdrawal certificate proof}
\end{center}

Withdrawal certificate proof is a SNARK proof that validates compliance of the certificate with a set of predefined rules. 

As it is defined in [Def. \ref{def:snark}], a SNARK is a proving system. Its particular instantiation is specified by a set of arithmetic constraints defining the verification rules. Each sidechain specifies its own set of constraints for the withdrawal certificate SNARK, thus establishing its own rules.

The basic interface for the SNARK prover and verifier is the following:
\[ proof \leftarrow Prove(pk_{WCert},\ public\_input,\ witness),\]
\vspace*{-6mm}
\[ true/false \leftarrow Verify(vk_{WCert},\ public\_input,\ proof).\]

A particular instantiation of the SNARK proving system is determined by a pair of keys -- proving key $pk_{WCert}$ and verifying key $vk_{WCert}$. Verifying key is registered upon sidechain creation and cannot be changed during the SC lifetime. It completely defines the rules of the withdrawal certificate validation (including the semantics of the public input and witness for the prover and verifier).

Recall from [Def. \ref{def:wcert}] that $public\_input$ for the WCert SNARK is comprised of two parts:
\[ public\_input\ \stackrel{\mathrm{def}}{=}\ (wcert\_sysdata,\ MH(proofdata)),\]
where $wcert\_sysdata$ is a set of arguments enforced directly by the mainchain\footnote{These arguments are explained in [\ref{sec:4_1_2_BackwardTransfers} \nameref{sec:4_1_2_BackwardTransfers}]}:
\[wcert\_sysdata\ \stackrel{\mathrm{def}}{=}\ (quality,MTHash(BTList),H(B_{len}^{i-1}),H(B_{len}^i));\]
and $proofdata$ is a set of arguments defined by the sidechain construction and passed along the withdrawal certificate. In the Latus sidechain, it is defined as follows:

\[proofdata\ \stackrel{\mathrm{def}}{=}\ (H(SB_{last}^{i}),H(state_{SB_{last}^i}[MST]),mst\_delta),\]
where
\begin{conditions}
    H(SB_{last}^i) & a hash of the last sidechain block in the epoch $i$ for which the certificate is created; \\

    H(state_{SB_{last}^i}[MST]) & a root hash of the MST tree derived after applying $SB_{last}^i$; note that by including $H(state_{SB_{last}^i}[MST])$ in $proofdata$, the withdrawal certificate commits to the updated sidechain state;\\

    mst\_delta & a bit vector of MST modifications; given that the MST is a fixed size Merkle tree, $mst\_delta$ is also a fixed-size bit vector where each bit represents a particular leaf in the tree; the bit is set to “1” if the MST leaf has been modified at least once during the epoch, otherwise it is “0” (see example in [Appendix \ref{sec:AppendixA}]).
\end{conditions}

$mst\_delta$ is used for proving that some $utxo$ has not been spent since some moment in the past (this is particularly useful for preventing data availability attacks as it allows creating mainchain managed withdrawals without knowing the current sidechain state). E.g., to prove this, one would need to provide a $utxo$ together with a Merkle proof of its inclusion in some $state_{SB_{last}^k}[MST]$ committed in one of the previous certificates and a list of $mst\_delta$’s from the following certificates where the corresponding bit has not been triggered to “1”. 
\\~\\
In the Latus sidechain construction, a withdrawal certificate proof enforces the following rules:

\begin{protocolframe}{\textbf{WCert SNARK Statement}}
\begin{itemize}
    \item $SB_{last}^i$ is the last block of the withdrawal epoch $i$ for which the certificate is created.
    \item $SB_{last}^i$ is connected to the $SB_{last}^{i-1}$ from the previous withdrawal certificate by a valid chain of blocks.
    \item $H(state_{SB_{last}^i}[MST])$ is a valid root of the MST for $state_{SB_{last}^i}$.
    \item Assuming that after applying the block $SB_{last}^{i-1}$ the sidechain state is $state_{SB_{last}^{i-1}}$ and after the block $SB_{last}^i$ the state is $state_{SB_{last}^i}$ , the proof verifies correct transition from $state_{SB_{last}^{i-1}}$ to $state_{SB_{last}^i}$ which means that all transactions from the subchain $[SB_0^i,...,SB_{last}^i]$ are correctly processed according to the rules from [\ref{sec:5_3_TransactionalModel}~\nameref{sec:5_3_TransactionalModel}].
    \item MC blocks from range $[B_0^i,...,B_{last}^i]$ are referenced from the sidechain blocks $[SB_0^i,...,SB_{last}^i]$ (this also implies that all SC-related transactions from these blocks have been processed).
    \item $BTList$ is a valid list of backward transfers that corresponds to $state_{SB_{last}^i}[backward\_transfers]$.
    \item $quality$ parameter is the height of the block $SB_{last}^i$.
    \item $mst\_delta$ is a bit vector that reflects changes in MST between $state_{SB_{last}^{i-1}}[MST]$ and $state_{SB_{last}^i}[MST]$.
\end{itemize}
\end{protocolframe}

In general, a withdrawal certificate proof validates correct transition for a range of blocks that belongs to the withdrawal epoch and that this range is adjacent to the range committed in the previous withdrawal certificate. This includes proving the correctness of backward transfers.

Given that all state transitions are proved, it becomes infeasible to create a malicious backward transfer (without creating a corresponding transaction in the sidechain), and it is infeasible to create new coins on the sidechains without real forward transfers.

\subsubsubsection{Backward Transfer Request} \label{sec:5_5_3_2_BTR}

In [\ref{sec:5_3_4_BTRTx} \nameref{sec:5_3_4_BTRTx}], we have already discussed how BTRs are submitted and processed in the sidechain. Here, we only provide details about the SNARK proof included in a BTR.

Recall that the BTR structure has been defined as follows [Def. \ref{def:btr}]:
\[BTR\ \stackrel{\mathrm{def}}{=}\ (ledgerId,\ receiver,\ amount,\ nullifier,\ proofdata,\ proof).\]

$proofdata$ is defined by the Latus construction as:
\[proofdata=\{utxo\},\]
where $utxo$ is an unspent output that holds coins that a user wants to withdraw. The basic idea is that the $proof$ should validate the user’s right to withdraw this $utxo$ and that this $utxo$ is present in the sidechain state MST committed by the last withdrawal certificate included in the mainchain.

Note that the BTR SNARK $proof$ is validated by the mainchain upon BTR submission. Even though it verifies that the withdrawn $utxo$ has been present in the last committed SC state, it cannot guarantee that it will remain valid at the moment BTR will be synchronized to the sidechain. This proof serves more like a pre-validation for the BTR in the mainchain to impede submission of wittingly invalid requests. 

The basic interface for the SNARK prover and verifier is the following [Def. \ref{def:btr}]:
\[ proof \leftarrow Prove(pk_{BTR},\ public\_input,\ witness),\]
\vspace*{-6mm}
\[ true/false \leftarrow Verify(vk_{BTR},\ public\_input,\ proof).\]

The verifying key $vk_{BTR}$ is registered upon sidechain creation. It defines the rules of the BTR validation (including the semantics of the public input and witness for the prover and verifier).

The $public\_input$ comprises two parts [Def. \ref{def:btr}]:
\[ public\_input\ \stackrel{\mathrm{def}}{=}\ (btr\_sysdata,\ MH(proofdata)),\]
\vspace*{-6mm}
\[ btr\_sysdata\ \stackrel{\mathrm{def}}{=}\ (H(B_w),\ nullifier,\ receiver,\ amount)),\]
where $H(B_w)$ is the hash of the MC block with the latest withdrawal certificate (at the moment when BTR is included in the mainchain), $receiver$, $amount$, and $nullifier$ are taken from the BTR itself. Note that $btr\_sysdata$ is enforced by the mainchain so its parameters cannot be manipulated by the BTR issuer.

A BTR proof enforces the following rules:
\begin{protocolframe}{\textbf{BTR SNARK Statement}}
\begin{itemize}
    \item $H(B_w)$ is the hash of the mainchain block where the last certificate $WCert_w$ has been submitted for this sidechain.
    \item $utxo\in{state_w[MST]}$, where $state_w[MST]$ has been committed in $WCert_w$.
    \item The BTR issuer has rights to spend this $utxo$ (i.e., possesses the corresponding private key).
    \item $amount$ is equal to the  $utxo.amount$.
    \item $nullifier$ is the hash of the $utxo$.
    \item $receiver$ is the address of the receiver in the mainchain.
\end{itemize}
\end{protocolframe}

\subsubsubsection{Ceased Sidechain Withdrawal} \label{sec:5_5_3_3_CSW}

CSWs are used to allow sidechain stakeholders to withdraw coins from a ceased sidechain.

As it has been defined in [Def. \ref{def:CSW}], a ceased sidechain withdrawal is submitted to the mainchain as a special transaction and performs a direct payment in the mainchain. Recall the basic structure of the ceased sidechain withdrawal that is defined by the mainchain [Def. \ref{def:CSW}]:
\[CSW\ \stackrel{\mathrm{def}}{=}\ (ledgerId,\ receiver,\ amount,\ nullifier,\ proofdata,\ proof).\]

The main prerequisite for CSW validity is the existence of the claimed coins in the sidechain state committed by the last withdrawal certificate. A sidechain user should point to the specific unspent output from $state_{SB_{last}^i}[MST]$ and authorize its spending. Basically, it is the same SNARK that is used for the BTR [\ref{sec:5_5_3_2_BTR} \nameref{sec:5_5_3_2_BTR}]; the difference is that now it authorizes direct payment in the mainchain, whereas in the BTR, it is essentially a pre-validation.

We will not dive deeply into the SNARK construction for the CSW as technically it is completely the same as for the BTR.

In general, the CSW proof validates that a submitter owns the utxo with a particular amount of coins at the moment of the sidechain halt. Also, it enforces a nullifier which is a unique identifier of the withdrawn utxo. Nullifiers are tracked by the mainchain to prevent withdrawal of the same coins twice.

\section{Conclusions}

The concept of sidechains has been acknowledged as an appealing solution for enhancing existing blockchain systems. It allows creating platforms and applications that are bound to the mainchain without imposing significant burden. Yet, we have not seen wide adoption of this concept. We believe that the value of sidechains as a scalability solution is underestimated and seek to develop this area.

In this paper, we introduced Zendoo, a universal construction for blockchain systems that enables the creation and communication with different sidechains without knowing their internal structure. We also provided a specific sidechain construction, Latus, that leverages zk-SNARK techniques to establish decentralized and verifiable cross-chain transfers.

We consider this as a research paper whose subject is still under ongoing research. In future publications, we plan to uncover more details about specific components and properties of the proposed sidechain construction.

\section{Acknowledgments}

We would like to express great appreciation to Maurizio Binello and Andrey Sobol for participating in technical discussions.

We would also like to thank Rob Viglione, Daniele Di Benedetto, Marcelo Kaihara, Luca Cermelli, and Lyudmila Kovalchuk for reviewing and providing valuable comments.
\bibliographystyle{plain}
\bibliography{references}
\begin{appendices}
\section{MST Delta} \label{sec:AppendixA}

Here, we provide an example of how the $mst\_delta$ value from a withdrawal certificate [\ref{sec:5_5_2_1_WCert} \nameref{sec:5_5_2_1_WCert}] is calculated and give some explanations on why it is needed. Note that this relates only to the Latus sidechain construction.

In general, $mst\_delta$ shows which leaves have been changed between two Merkle state trees [\ref{sec:5_2_Accounting} \nameref{sec:5_2_Accounting}] $MST_i$ and $MST_j$, $i < j$ (e.g., in case of a withdrawal certificate, these are MSTs committed by the previous certificate and the current one which shows how the system state changed during the epoch). $mst\_delta$ is a bit vector that shows what leaves of the $MST_i$  have been changed in $MST_j$.

\begin{figure}[htbp]
	\centering
	\includegraphics[trim={2cm 12.6cm 7cm 2cm}, clip,width=0.95\columnwidth] {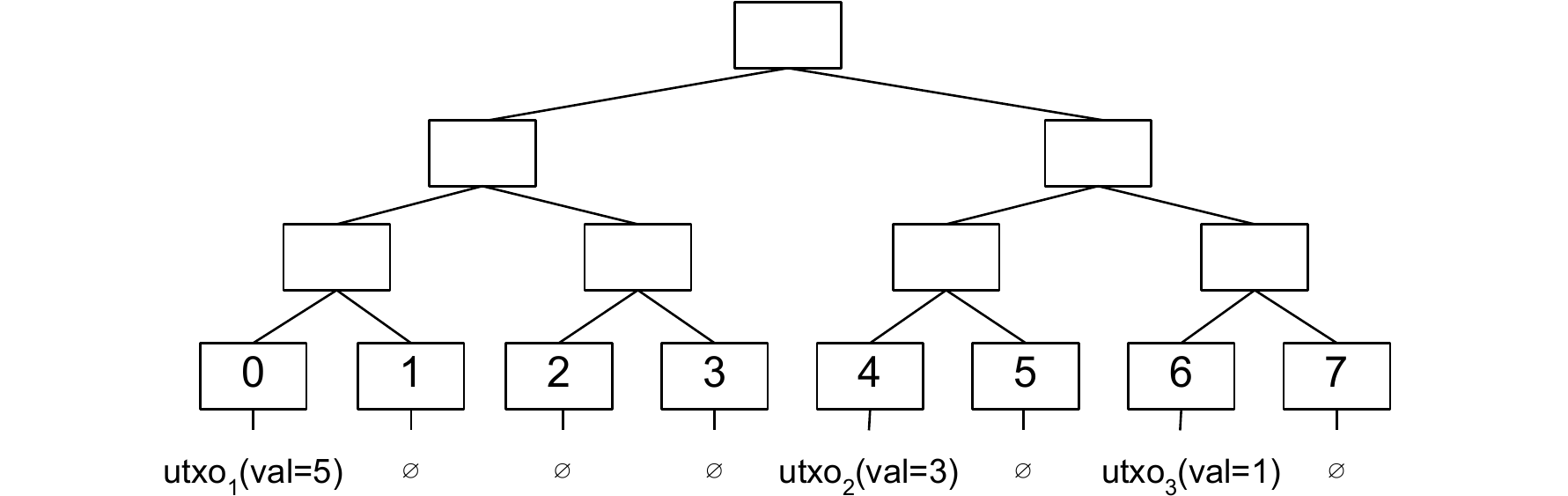}
	\caption{Merkle state tree $MST_0$.}
	\label{fig:A_1}
\end{figure}

Let us consider the MST of depth $D_{MST}=3$ which has an initial state $MST_0$ (see Fig. \ref{fig:A_1}).

The MST can contain up to eight unspent outputs (equal to the number of leaves). At the moment $MST_0$, the tree contains three UTXOs $\{utxo_1, utxo_2, utxo_3\}$ which are assigned to leaf nodes {0, 4, 6} correspondingly.

\vbox{%
Let us assume that we have two transactions $tx_1$ and $tx_2$ such that:
\begin{protocolframe}{}
\begin{lstlisting}
tx1 = {
	inputs: {utxo1}
	outputs: {utxo4(val=2), utxo5(val=3)}
}
tx2 = {
	inputs: {utxo4}
	outputs: {utxo6(val=2)}
}
\end{lstlisting}
\end{protocolframe}
}
Assuming that $MST\_Position(utxo_4)=1$, $MST\_Position(utxo_5)=2$, and\\ $MST\_Position(utxo_6)=7$, applying transactions $tx_1$ and $tx_2$ to the state $MST_0$  will provide the following $MST_1$:
\begin{figure}[htbp]
	\centering
	\includegraphics[trim={2cm 12.6cm 7cm 2cm}, clip,width=0.95\columnwidth] {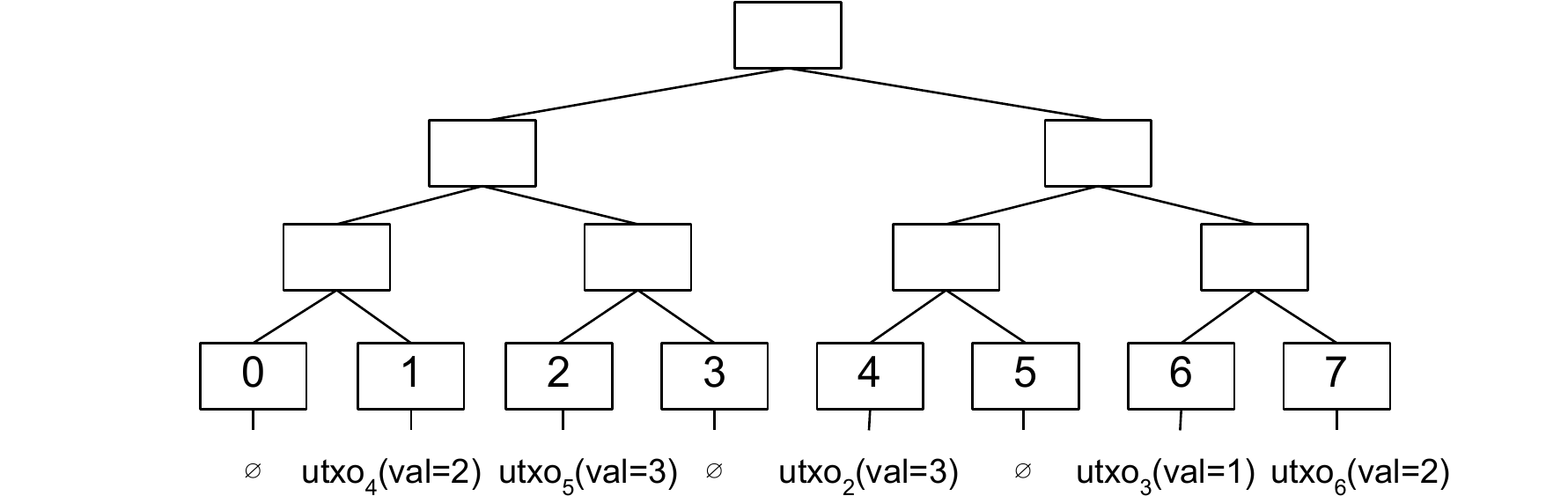}
	\caption{Merkle state tree $MST_1$.}
	\label{fig:A_2}
\end{figure}

It can be seen that during the transition from $MST_0$ to $MST_1$, the leaves {0, 1, 2, 7} have been modified. Thus, the $msd\_delta$ reflects these modifications in the bit vector:
 \[mst\_delta=(11100001),\]
where each bit represents whether a corresponding leaf node has been modified.
\\~\\
Having $mst\_delta$ in each withdrawal certificate allows to prove that some $utxo_a$ is contained in $MST_k$  committed by the latest certificate, by providing proof of inclusion in some $MST_t$, $t < k$, committed by the certificate in the past, and verifying that the bit $MST\_Position(utxo_a)$ is zero for all $mst\_delta$’s on the way from $MST_t$ to $MST_k$.

This feature is of great value for circumventing data availability attacks, e.g., when a compromised sidechain (where the majority of stakeholders is adversarial) submits a withdrawal certificate to the mainchain that commits to some $MST_k$ while not revealing to the public the $MST_k$ tree itself. Having $mst\_delta$ in place, a user will be able to create proof of utxo ownership by using some previous $MST_k$. This mechanism is used for proving utxo ownership in mainchain managed withdrawals in the Latus sidechain construction ([\ref{sec:5_5_3_2_BTR} \nameref{sec:5_5_3_2_BTR}], [\ref{sec:5_5_3_3_CSW} \nameref{sec:5_5_3_3_CSW}]).

\end{appendices}
\end{document}